\RequirePackage[2020/08/10]{latexrelease}
\documentclass[aps,pra,reprint,superscriptaddress,longbibliography]{revtex4-2}

\usepackage{physics}
\usepackage{amsmath}
\usepackage{amssymb}
\usepackage{graphicx}
\usepackage{float}
\usepackage{mathrsfs}
\usepackage{mathtools}
\usepackage{extarrows}
\usepackage[margin=0.7in]{geometry}
\graphicspath{{PNG/}}
\usepackage{xcolor}

\begin{document}
	
	\title{Focusing Atom Laser Beams}
	
	\author{R.~Richberg}
	
	\email{formerly known as Amir M. Kordbacheh}
	
	\affiliation{ 
		Department of Quantum Science, Research School of Physics, The Australian National University, Canberra, ACT 2601, Australia
	}
	
	\affiliation{ 
		School of Physics, University of Melbourne, Melbourne, 3010, Australia
	}%

	\author{A.~M.~Martin}%
	
	\affiliation{ 
		School of Physics, University of Melbourne, Melbourne, 3010, Australia
	}%

	\date{\today}

	\begin{abstract}
		
		We theoretically study the focusing of a quasi-continuous atom laser beam of rubidium-85 ($^{85}$Rb). A two-sate model analysis based on the Gross-Pitaevskii equation is used which comprises the effects of two-body atom-atom interactions and three-body recombination losses. Utilizing optical focusing potentials such as harmonic potentials, the essential factors such as the width, peak density and atom loss rate of the focused atom laser beam profile are investigated. Our analysis predicts that using an atom laser offers a dramatic improvement in resolution of up to $8$ nm.
		
	\end{abstract}
	
	\maketitle

	\section{Introduction}

	The development of laser cooling and trapping of atoms \cite{I1, I2, I3, I4} has enabled the observation of coherent, wave-like properties of neutral atoms. Atomic matter-waves can be manipulated with high precision using optical light fields for advanced atomic physics experiments. Not only has this innovation brought outstanding applications in quantum computing \cite{I11, I12}, quantum entanglement \cite{I14}, nonlinear phenomena \cite{I15} and quantum turbulence \cite{I16}, but also it has been utilized in the precision measurement of inertial forces using atom interferometries, such as state-of-the-art measurements of accelerations \cite{I18, I19}, rotations \cite{I20, I21}, gravity \cite{I22, I23, I24}, gravity gradients \cite{I25, I26} and magnetic fields \cite{I27}. 
	
	Nonetheless, the controlling of matter-waves using optical potentials opens doors to the requirements of prospective experiments in atom lithography and nano-fabrication; using a bright, coherent, almost perfectly collimated (extremely low divergence) ultra-cold atomic beam derived from a Bose-Einstein condensate (BEC) \cite{3_1} in the production of nano-scale devices is an emerging area providing superior structure linewidths. This is deemed to break the limitations of previous atom lithography studies with traditional oven-based sources \cite{p1, p2, p49, p50, p51,p52}. In almost all these efforts, the fabricated structures suffer from high transverse temperatures leading to divergence of the incoming beam causing a broadening on the deposited surface. In addition, the chromatic aberration resulting from a broad longitudinal velocity distribution while focusing through an optical potential can, in turn, blur the resolution. However, atom lithography using a low-divergent atom laser beam \cite{5_1,5_32} has the capability to enhance the resolution of the deposition process dramatically as it provides a high flux and a coherently mono-chromatic collimated beam of atoms. It is worth noting that atom deposition using propagating BECs has recently been theoretically considered \cite{Ri_3,Ri_4,Ri_1,Ri_2} leading to predicted resolutions of $20$ and $9$ nm via various external focusing potential configurations. 
	
	The examination of the evolution of an atom laser is well established. Riou {\it et al}. \cite{AL_1} introduced the \textit{ABCD} matrices formalism as analytical tools to estimate the non-interacting atom laser dynamics at different stages of free propagation. This analysis has been extended perturbatively beyond the linear regime of propagation to account for interacting atomic clouds \cite{AL_2}. In 2009, utilizing linear \textit{ABCD} matrices, Impens investigated the propagation of a paraxial interacting atomic beam in time-dependent, cylindrical, and quadratic potentials \cite{AL_3}. 
	
	The process of extracting a small fraction of the condensate in the form of atomic coherent beam without disturbing the rest of the condensate is known as an outcoupling technique \cite{5_6, 5_13}. There are various outcoupling approaches used from which RF (Radio Frequency) \cite{5_6} and stimulated Raman outcoupling \cite{5_7} are the well-studied methods. In a RF outcoupled atom laser \cite{5_6, 5_9, 5_10}, a low intensity RF field is transmitted through the BEC causing some of cloud atoms to flip their magnetic substate (Zeeman state) from the trapped state $m_F=-1$ to the untrapped state $m_F=0$ exiting the BEC. However, in stimulated Raman process \cite{5_7, 5_11, 5_12}, atoms are transferred from the trapping state to non-trapping state using a Raman transition by crossing through the condensate two counter-propagating tunable lasers of frequencies of $\omega_1$, $\omega_2$ and wavevectors of $\mathbf{k_1}$, $\mathbf{k_2}$; atoms can undergo Raman transitions from the trapped state to the untrapped state by exchanging photons between two lasers. It was understood in \cite{AL_4} that the angular divergence of a quasi-continuous, RF-outcoupled and free-falling atom laser is dominated by the condensate-laser interaction during the outcoupling process. Furthermore, it was demonstrated by M. Jeppesen {\it et al}. \cite{5_18} that using an optical Raman transition rather than RF field can eliminate the diverging lens effect that the condensate has on the outcoupled atoms leading to the improvement of the beam quality of the atom laser.

	In this paper, we use a two-state model \cite{5_31} based on the Gross-Pitaevskii equation (GPE) \cite{3_8, 3_9} formulating, specifically, the focusing dynamics of $^{85}$Rb quasi-continuous atom laser state. We assume that a BEC is confined inside an optical dipole trap. Utilizing a Bragg diffraction outcoupling technique \cite{5_8}, a fraction of atoms is scattered from the BEC forming an atom laser state whilst carrying a momentum kick. While the model considers two-body inter-atomic interactions dominated by \textit{s}-wave scattering \cite{10,11,12,13}, it is developed to account for the three-body recombination losses \cite{15,16,17} as well as the external focusing potential field. Our model exploits a harmonic optical focusing potential, with optimal focal parameters achieved via a classical trajectories approach \cite{20}. Then, we examine the focusing of atom laser beams in the context of a focusing potential, two-body inter-atomic interactions as well as three-body recombination losses. The characteristic factors of deposited structures such as the resolution and peak density are ultimately estimated over a wide range of two-body interactions, which can be controlled via a Feshbach resonance \cite{29} within the beam. The influence of focusing potential geometries, magnitude of Rabi outcoupling frequency \cite{5_11,5_21}, momentum kicks applied to the beam is the matter of consideration in the numerical simulations. Within this model, we obtain the requirements of producing a structure's resolution of about $8$ nm for a focused atom laser beam.

	\section{Bragg Outcoupling from a BEC}
	\label{subsec:1}
	
		\begin{figure}[b]
		\centering
		\includegraphics[width=7cm, height=3cm,angle=0] {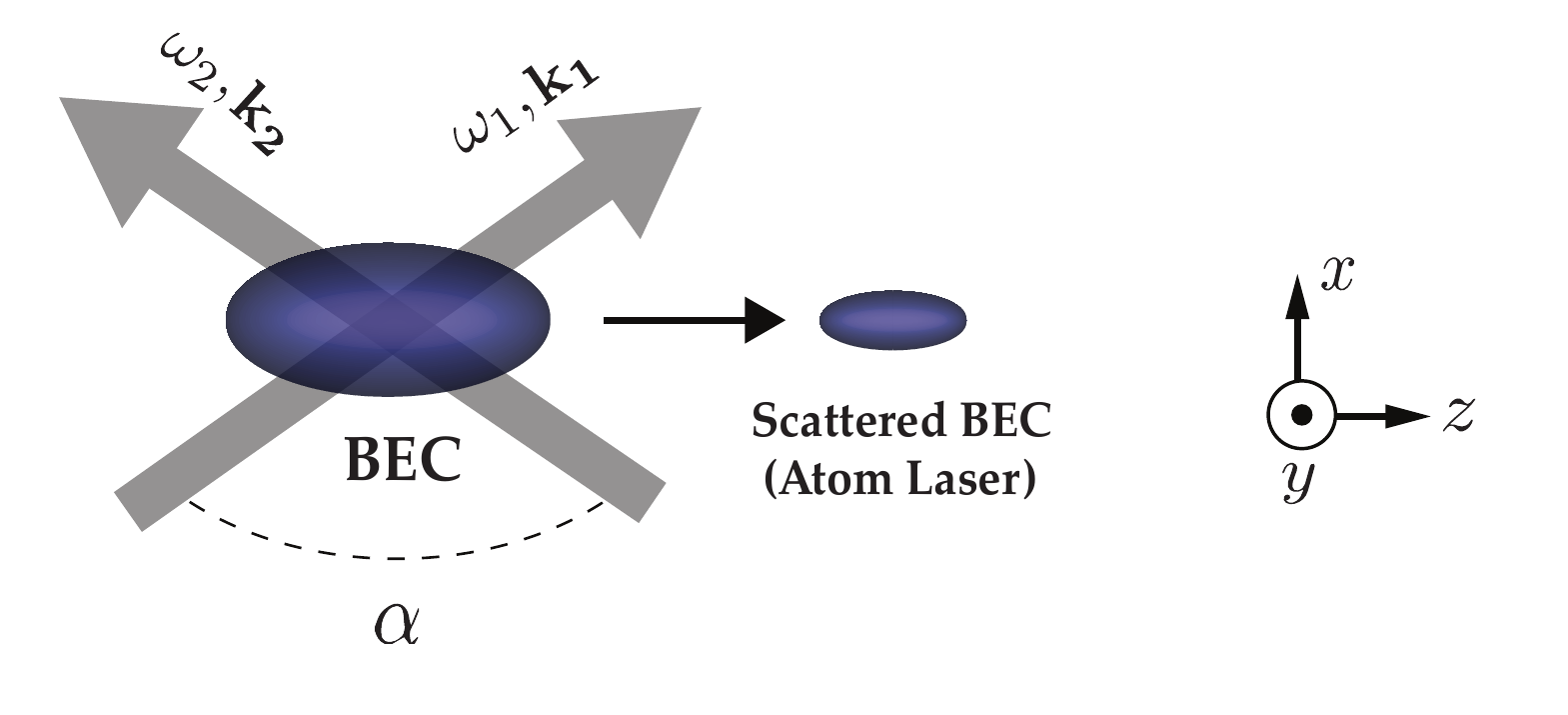}
		\caption{Schematic of the arrangement of the laser fields in Bragg outcoupling event.}
		\label{f509}
	\end{figure}
	
		\begin{figure}[t!]
		\centering
		\includegraphics[width=9cm, height=8cm,angle=0] {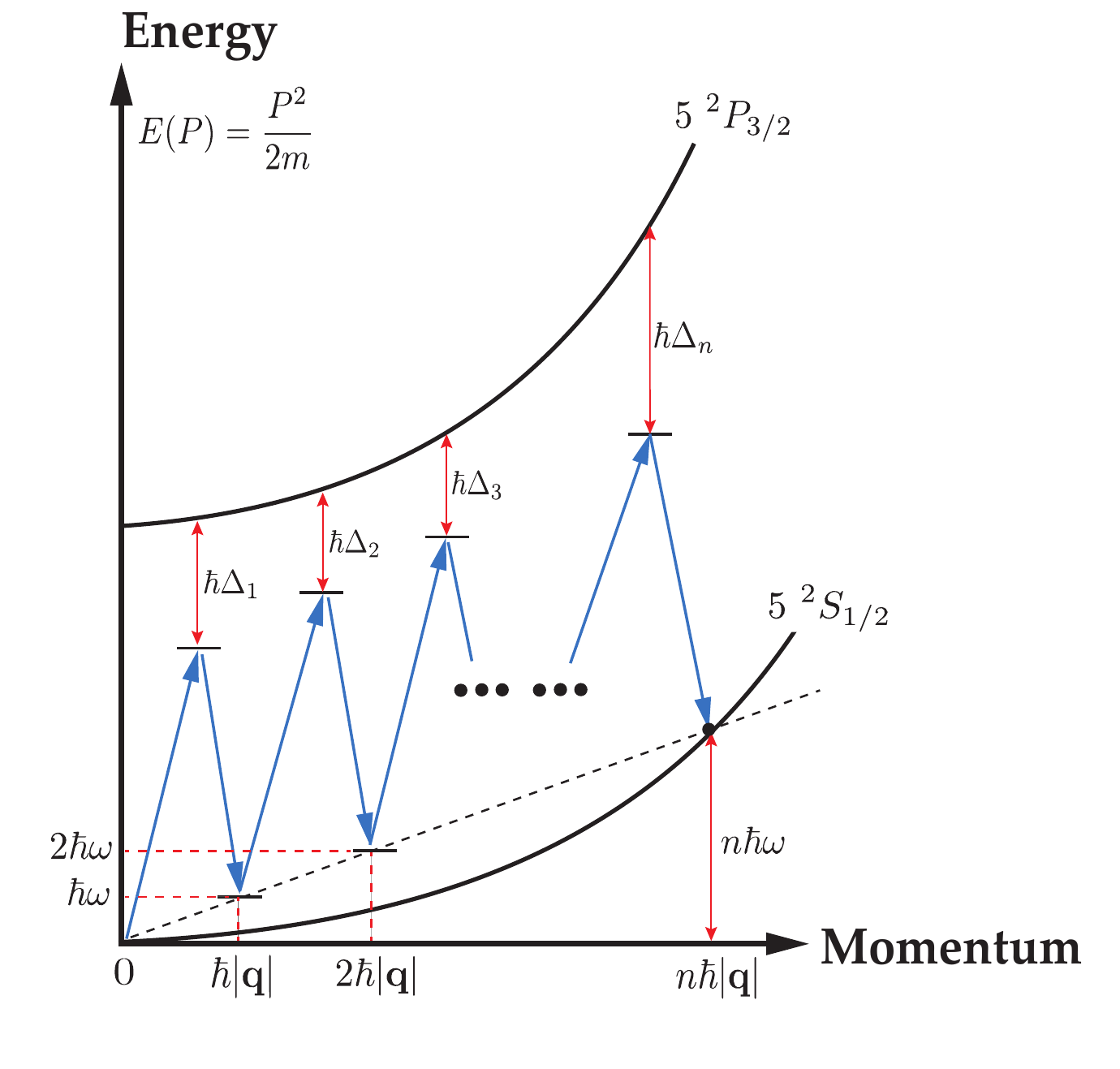}
		\caption{Schematic illustration of $n$ two-photon Bragg scattering process. The upper and lower solid curves ($5~^2P_{3/2}$ and $5~^2S_{1/2}$ states) show the internal states in terms of energy and momentum for $^{87}$Rb. The single black dot on the energy-momentum parabola indicates the final momentum state arising from the total energy conservation corresponding to the Bragg resonant transition. The magnitudes of energy and momentum transferred to atoms in each two-photon event are dependent on the value of resultant wavenumber, $q$, and consequently the angle between lasers, $\alpha$. For every two-photon process, the associated detuning from resonance is depicted by $\Delta_i$ where $i=1, ..., n$.}
		\label{f510}
	\end{figure}

	One way of applying momentum to a condensate trapped inside an optical dipole trap is via a two-photon scattering process supplied by two laser fields. In this scenario, atoms inside the condensate absorb and re-emit photons between two fields and recoil with a magnitude equal to the momentum difference of two photons. This process is known as the Bragg scattering of a BEC in which the scattered condensate (also referred to as the atom laser) will remain in the same internal state \cite{5_25}. Bragg diffraction is an efficient outcoupling procedure of producing an atom laser \cite{5_6}, and the main benefit in this process is producing a directed output beam with an excellent coherence \cite{5_26} while carrying a large momentum transfer. As illustrated in Fig~\ref{f509}, two laser fields with frequencies of $\omega_1$ and $\omega_2$, and the wavevectors of $\mathbf{k}_1$ and $\mathbf{k}_2$ at the angle of $\alpha$ transfer a momentum to the stationary condensate and kick a fraction of atoms out of the condensate into a scattered state. The significant difference between the Raman outcoupling and Bragg scattering (Bragg outcoupling) techniques when using $^{87}$Rb or $^{85}$Rb is that in the former, the Raman transitions cause some of atoms to be located in $|5\ ^2S_{1/2}, F=1, m_F=0\rangle$ whereas in the latter, the Bragg transitions transfer the outcoupled atoms to the $|5\ ^2S_{1/2}, F=1\rangle$, which means that the vibrational levels for $F=1$ are degenerate in the absence of a magnetic field. On the other hand, the two outcoupling schemes are analogous in imparting a momentum kick to the outcoupled atoms. It is essential to select the frequency difference $\omega=\omega_1-\omega_2$, which brings the scattered condensate resonantly to a particular momentum state. In other words, $\omega$ needs to be tuned such that the scattered condensate momentum state ends up on the atomic energy-momentum parabola as depicted in Fig~\ref{f510}, where the Bragg transitions for the case of $n$ two-photon transitions bringing the scattered BEC to its final momentum state is displayed. To derive the resonance condition, we assume a sequence of $n$ two-photons are incident to the condensate at rest inside a dipole trap. In each two-photon transition, the momentum and energy delivered to the condensate by the fields are given by $\hbar(\mathbf{k_1}-\mathbf{k_2})$ and $\hbar \omega$ respectively. By the $n$th process, the recoil condensate (scattered condensate) in its final momentum state carries the total momentum and energy of
\begin{equation}
		\mathbf{p}_{\text{recoil}}=n\hbar(\mathbf{k_1}-\mathbf{k_2})=n\hbar \mathbf{q},
		\label{512}
	\end{equation}
	\begin{equation}
		E=n\hbar\omega,
		\label{513}
	\end{equation}
	where $|\mathbf{q}|=2k \sin(\alpha/2)$ for $|\mathbf{k}_1|=|\mathbf{k}_2|\approx k$, where $k=2\pi/\lambda$ and $\lambda$ is the wavelength of the light. The recoil action appears as the kinetic energy of the scattered BEC
	\begin{equation}
		E_{\text{recoil}}=\frac{p_{\text{recoil}}^2}{2m},
		\label{514}
	\end{equation}
	where $m$ is the atomic mass. According to the conservation of energy and momentum, by equating the energy provided by the fields in the $n$ two-photon processes, Eq. (\ref{514}), with the recoil energy, Eq. (\ref{513}), one can obtain the Bragg resonance condition \cite{5_25}
	\begin{equation}
		\omega_{\text{res}}=\frac{n\hbar q^2}{2m}.
		\label{515}
	\end{equation} 
	The resonance condition for the case of $n$ two-photon is shown in Fig~\ref{f510}. The first $n-1$ transitions do not end up on the energy-momentum parabola curve, therefore are not satisfied energetically by the resonance condition,  Eq. (\ref{515}). However, the last two-photon transition (the $n$th transition) allows energy and momentum conservation satisfying the resonance condition, and hence the transition $|0\rangle\longrightarrow|n\hbar\mathbf{q}\rangle$ in momentum space occurs.

	\section{Two-State Model for the BEC and Atom Laser}
	\label{subsec:2}
	
	The incident Bragg pulses to the BEC create a standing wave potential encapsulating the condensate. The  Gross-Pitaevskii equation for the ground state of a condensate when exposed to an optical dipole potential produced by Bragg pulses is estimated \cite{5_28} as 
	
	\begin{equation}
		\begin{split}
			i\hbar\frac{\partial\psi(\mathbf{r}, t)}{\partial t}=-\frac{\hbar^2}{2m}\nabla^2\psi(\mathbf{r}, t)+V_{T}(\mathbf{r})\psi(\mathbf{r}, t)~~~~~~~~~\\
		+V_{\text{opt}}(\mathbf{r}, t)\psi(\mathbf{r}, t)
			+u|\psi(\mathbf{r}, t)|^2\psi(\mathbf{r}, t),
			\label{544}
		\end{split}
	\end{equation} 
	where $\hbar$ and $m$ are, respectively, Planck's constant and the atomic mass for rubidium-85, and $V_{T}(\mathbf{r})=m\big(\omega_{x}^2x^2+\omega_{y}^2y^2+\omega_{z}^2(z)^2\big)/2$ is the harmonic trapping potential. The mean-field potential term is described by $u|\psi(\mathbf{r},t)|^{2}$, where $u=\frac{4\pi\hbar^2a_s}{m}$ is the inter-atomic interaction strength, $|\psi(\mathbf{r},t)|^{2}$ is the atomic density, and $a_s$ is the $s$-wave scattering length \cite{10,11,12,13}. The optical potential for the ground state in an interaction picture is represented by
	\begin{equation}
		V_{\text{opt}}(\mathbf{r}, t)=2\hbar \Omega(t)\cos(\mathbf{q}\cdot\mathbf{r}-\omega t),
		\label{545}
	\end{equation} 
	where $\Omega(t)$ represents the two-photon Rabi frequency.
	
	Since the magnitude of the recoil momentum of the Bragg scattering is much larger than the momentum width of the initial condensate, the scattered states accompanied by different momentum values are distinguishable in momentum space, $\phi_n({\mathbf{k}}, t)$, and consequently in position space, $\psi_n({\mathbf{r}}, t)$ \cite{5_33}. According to the slow-varying envelope approximation \cite{5_29}, the total wavefunction of the system, $\psi({\mathbf{r}, t})$, consists of $n$-wavepackets moving with different central momenta, $\mathbf{p}_n=n\hbar\mathbf{q}$, can be written as
	\begin{equation}
		\psi({\mathbf{r}, t})=\sum_{n}\psi_n({\mathbf{r}, t})\exp\big[{in(\mathbf{q}\cdot\mathbf{r}-\omega t)}\big],
		\label{546}
	\end{equation} 
	where $\psi_0({\mathbf{r}, t})$ is the stationary ``mother'' BEC with a momentum wavepacket centered at zero, whereas $\psi_1({\mathbf{r}, t})$, $\psi_2({\mathbf{r}, t})$, ..., $\psi_n({\mathbf{r}, t})$ are the scattered wavepackets (atom laser states) with the momentum value of $|\mathbf{p}_1|=\hbar|\mathbf{q}|$, $|\mathbf{p}_2|=2\hbar|\mathbf{q}|$, ..., $|\mathbf{p}_n|=n\hbar|\mathbf{q}|$ in 2-photon, 4-photon, ..., 2$n$-photon processes respectively.
	
	The relation between the complete wavefunction of the system, $\psi({\mathbf{r}, t})$, and the partitioned wavefunctions, $\psi_n({\mathbf{r}, t})$ is achieved using a number of transforms. Firstly, a direct Fourier transform, $\mathscr{F}$, transfers the wavefunction, $\psi({\mathbf{r}, t})$, into momentum space forming the total momentum wavefunction of the system, $\phi({\mathbf{k}, t})$. Secondly, this becomes restricted to a specific domain in momentum space, $\mathscr{K}_n$, resulting in the partitioned momentum wavefunctions, $\phi_n({\mathbf{r}, t})$. Finally, an inverse Fourier transform, $\mathscr{F}^{-1}$, provides $\psi_n({\mathbf{r}, t})$ by transferring back the momentum space into position one. The procedure is mathematically represented by the following
	
	\begin{equation}
		\begin{split}
			\mathscr{F}^{-1}\Big(\mathscr{K}_n\Big(\mathscr{F}\psi({\mathbf{r}, t})\Big)=~~~~~~~~~~~~~~~~~~~~~~~~~~~~~~~~~~~~
			\\ 
			\frac{1}{(2\pi)^3}\int_{\mathscr{K}_n}d\mathbf{k}\ e^{i\mathbf{k}\cdot\mathbf{r}}\int d\mathbf{r}' e^{-i\mathbf{k}\cdot\mathbf{r'}}\psi({\mathbf{r'}, t}).
			\label{548}
		\end{split}
	\end{equation}
	Using Eqs.(\ref{546}) and (\ref{548}), the serial operators can be acted on both sides of Eqs.(\ref{544}) resulting in the GPE for the partitioned position states as
	\begin{widetext}
		\begin{equation}
			i\hbar\frac{\partial\psi_n(\mathbf{r}, t)}{\partial t}=\Big(-\frac{\hbar^2}{2m}\nabla^2+V_{T}(\mathbf{r})+n\hbar \omega_{\text{res}}+u|\psi(\mathbf{r}, t)|^2\Big)\psi_n(\mathbf{r}, t)
			+\hbar \Omega(t)\Big(e^{i(\mathbf{q}\cdot\mathbf{r}-\omega t)}\psi_{n+1}+e^{-i(\mathbf{q}\cdot\mathbf{r}-\omega t)}\psi_{n-1}\Big).
			\label{564}
		\end{equation}
	\end{widetext}
	Considering the case of Bragg scattering via two-photon ($n=1$), and choosing $\omega$ to satisfy the Bragg resonance condition, the coupling between $|\psi_0\rangle\leftrightarrow|\psi_1\rangle$ (or the momentum transition $|0\rangle\leftrightarrow|\hbar\mathbf{q}\rangle$) can be held while all couplings to other $|\psi_n\rangle$ are neglected. In this case Eq.(\ref{564}) is split into a set of two equations for $n=0$ (corresponding to zero momentum) and $n=1$ (corresponding to the momentum value of $\hbar|\mathbf{q}|$) which describe the stationary condensate and atom laser states respectively. In this instance, using Eq.(\ref{546}), the non-linear term in Eq.(\ref{564}) can be expanded to
	\begin{equation}
		\psi_n|\psi|^2=\psi_n\Big(\psi_0+\psi_1e^{i(\mathbf{q}\cdot\mathbf{r}-\omega t)}\Big)\Big(\psi_0^*+\psi_1^*e^{-i(\mathbf{q}\cdot\mathbf{r}-\omega t)}\Big),
		\label{565}
	\end{equation}
	which for the case of $n=0$ and $n=1$ reduces to
	\begin{equation}
		\psi_0|\psi|^2= \Big(|\psi_0|^2+|\psi_1|^2\Big)\psi_0;
		\label{566}
	\end{equation}
	\begin{equation}
		\psi_1|\psi|^2= \Big(|\psi_0|^2+|\psi_1|^2\Big)\psi_1,
		\label{567}
	\end{equation}
	in which the term $2\Re\big({\psi_0\psi_1^*e^{-i(\mathbf{q}\cdot\mathbf{r}-\omega t)}}\big)$ has been ignored by making the rotating-wave approximation (RWA). It is also worth considering the three-body losses component for the atom laser state indicated by $\psi_n|\psi|^4$. Similarly, using Eq.(\ref{546}), one can write
	\begin{equation}
		\psi_n|\psi|^4=\psi_n\Big(\psi_0+\psi_1e^{i(\mathbf{q}\cdot\mathbf{r}-\omega t)}\Big)^2\Big(\psi_0^*+\psi_1^*e^{-i(\mathbf{q}\cdot\mathbf{r}-\omega t)}\Big)^2.
		\label{568}
	\end{equation}
	Expanding Eq.(\ref{568}), and using the RWA, the forth order non-linear term for the case of $n=1$ can be written as
	\begin{equation}
		\psi_1|\psi|^4= \Big(|\psi_0|^4+|\psi_1|^4+4|\psi_0|^2|\psi_1|^2\Big)\psi_1.
		\label{569}
	\end{equation}
	We note that the $\psi_0|\psi|^4$, the BEC state, has been ignored due to the fact that no loss is considered within the BEC over the evolution time since, in experiments, the BEC is produced constantly with a fixed flux.
	
	Ultimately, considering Eqs.(\ref{566}), (\ref{567}) and (\ref{569}), and also extending the problem to the $n$ two-photon case, one can achieve the evolution equations for the BEC, $\psi_0$, and atom laser, $\psi_n$, states in an $n$ two-photon Bragg process as
	\begin{widetext}
		\begin{equation}
			i\hbar\frac{\partial\psi_0(\mathbf{r}, t)}{\partial t}=\Bigg(-\frac{\hbar^2}{2m}\nabla^2+V_{T}(\mathbf{r})+u(|\psi_0(\mathbf{r}, t)|^2+|\psi_n(\mathbf{r}, t)|^2)\Bigg)\psi_0(\mathbf{r}, t)\\
			+\hbar \Omega(t)e^{in(\mathbf{q}\cdot\mathbf{r}-\omega t)}\psi_n(\mathbf{r}, t);
			\label{570}
		\end{equation}
		\begin{equation}
			\begin{split}
				\begin{split}
					i\hbar\frac{\partial\psi_n(\mathbf{r}, t)}{\partial t}=\Bigg(-\frac{\hbar^2}{2m}\nabla^2+F_g+\frac{n^2\hbar^2q^2}{2m}+U_f(x, z)+u(|\psi_0(\mathbf{r}, t)|^2+|\psi_n(\mathbf{r}, t)|^2)\quad\quad\\
					-iK(|\psi_0|^4+|\psi_n|^4+4|\psi_0|^2|\psi_n|^2)\Bigg)\psi_n(\mathbf{r}, t)+\hbar \Omega(t)e^{-in(\mathbf{q}\cdot\mathbf{r}-\omega t)}\psi_0(\mathbf{r}, t),
					\label{573}
				\end{split}
			\end{split}
		\end{equation}
	\end{widetext}
	where $K$ indicates the three-body losses rate, which for $^{85}$Rb condensate is $4\times 10^{-41}$m$^6$s$^{-1}$ \cite{34, n6, II_3}, and $n\hbar\omega_{res}=\frac{n^2\hbar^2q^2}{2m}$ [see Eq.(\ref{515})]. Since the atom laser propagates freely under gravity along the $z$ direction, $\mathbf{g}=g\hat{\mathbf{z}}$, the trapping potential has been replaced by the gravitational potential, $F_g=m\mathbf{g}\cdot \mathbf{r}=mgz$, in the atom laser evolution equation [Eq.(\ref{573})] where $g=9.8$ m/s$^2$. The term $U_f(x, z)$, considered in Eq.(\ref{573}) represents the external focusing potential, which will be discussed in Section \ref{subsec:3}. Eqs.(\ref{570}), (\ref{573}) should be simultaneously solved to study the dynamics of the ``mother'' condensate and resultant atom laser.

	\section{The Focusing Potential}
	\label{subsec:3}
	
	We assume an optical focusing potential, $U_f(x, z)$, created by two counter-propagating laser lights resulting in a standing wave. The electric field induces a dipole moment in the atoms within the atomic beam. The interaction between the dipole moment and the electric field causes a dipole force \cite{37} with a gradient towards the nodes or anti-nodes of the laser light intensity. The resultant focusing potential \cite{19} is introduced as 
	\begin{equation}
		U_f(x, z)=\frac{\hbar \Delta}{2}\ln\Bigg(1+\frac{\gamma^2}{\gamma^2+4\Delta^2}\frac{I(x, z)}{I_s}\Bigg),
		\label{575}
	\end{equation}
	where $\Delta$ denotes the detuning of the laser frequency from the atomic resonance, $\gamma=38$ MHz the natural linewidth of the D$_2$ atomic transition of $^{85}$Rb and $I_s=1.67$ mW/cm$^2$ is the saturation intensity of the associated transition. The potential intensity profile is chosen to be harmonic shaped along the $x$ axis (the focusing direction) with a single node at $x=0$, which can be practically structured using a spatial light modulator \cite{38,39}. Assuming a Gaussian distribution along the $z$ axis (the direction of falling atom laser beam), this leads to $I(x,z)=I_0 \exp(-2z^2/\sigma_z^2)(k^2x^2)$, where $I_0$ is the maximum intensity of the harmonic profile, $\sigma_z$ is the radius of the beam at $1/e^2$ value of the maximum intensity, $k=2\pi/\lambda$ determines the strength of the harmonic focusing and $\lambda$ is the wavelength of the field. 
	
	By treating the atom dynamics as classical particle trajectories, the optimal laser power needed to focus the atoms at any desired spot along the focal axis ($z$-axis) can be determined \cite{20, Ri_1, Ri_3}
	\begin{equation}
		P_0=\xi\frac{\pi}{4}\frac{E_0}{\hbar\Delta}\frac{\gamma^2+4\Delta^2}{\gamma^2}\frac{I_s}{k^2},
		\label{e29}
	\end{equation}
	where $E_0$ is the initial kinetic energy of the atoms and $\xi$ is a dimensionless parameter. The power relates to the peak intensity via $I_0=8P_0/\pi\sigma_z^2$. A value of $\xi=5.37$, is determined by solving the classical equations of motion for atomic trajectories and optimally focuses the atoms onto the plane located at the center of the focusing potential along the focal $z$-axis. The selection of lower values of $\xi$ (smaller powers) leads to focusing on planes below the center of the focusing potential along the $z$-axis.
	
	\begin{figure}
		\centering
		\includegraphics[width=6cm, height=8.5cm,angle=0] {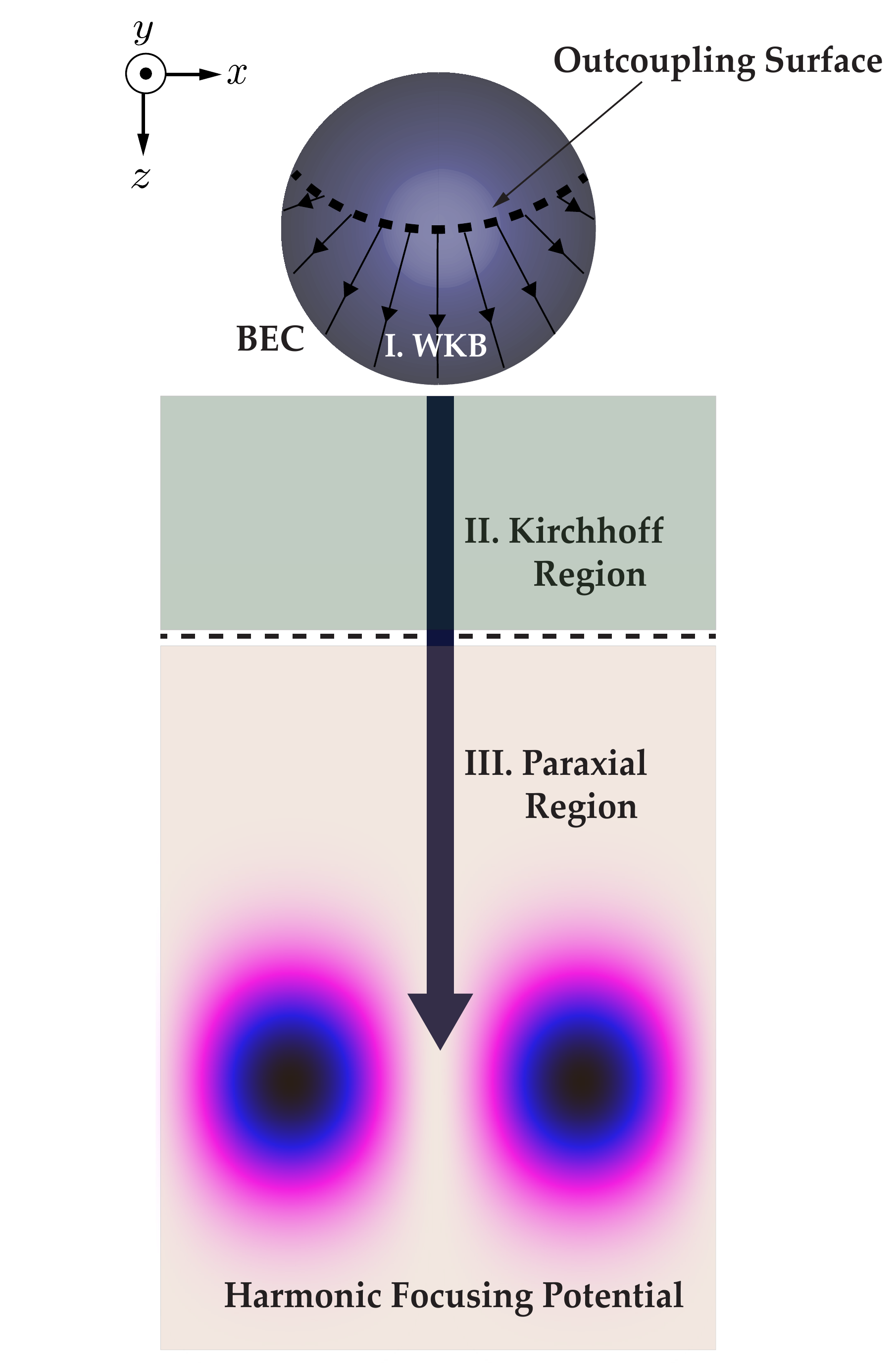}
		\caption{Schematic of the atom laser operation in three different regions: (I), (II) and (III) indicate the WKB inside the BEC, Kirchhoff and paraxial regimes respectively. The solid blue sphere shows the BEC while the solid dark blue arrow along the $z$-axis depicts the atom laser beam. The harmonic focusing potential has been illustrated from the top view ($x-z$ plane) by the two shaded oval areas at the bottom of the figure. The intensity of the potential reaches its maximum value at the black spots whereas the minimum intensity occurs between the two peaks.}
		\label{f511}
	\end{figure}
	
	\section{Results and Discussion}
	\label{subsec:4}
	
	The dynamics of an atom laser beam can be specifically analyzed in three individual regions \cite{5_13} shown in Fig~\ref{f511}: The WKB zone is the area inside the BEC between the outcoupling surface \cite{5_9} and bottom of the condensate. The early stages of propagation immediately below the bottom of the BEC is called the Fresnel-Kirchhoff zone, and the paraxial zone is where the atom laser has traveled sufficiently below the condensate. We utilize the two-state model described in section~\ref{subsec:2} [see Eqs.(\ref{570}), (\ref{573})], to numerically acquire the atom laser wavefunction over the entire path considering the two-body interactions and three-body losses. 
	
	In the following, the evolution of a quasi-continuous atom laser in both absence and presence of a focusing potential is discussed.


	\subsection{Atom Laser Properties Without Focusing}
	\label{subsec:5}

	\begin{figure*}
		\centering
		\includegraphics[width=14cm, height=21cm,angle=0] {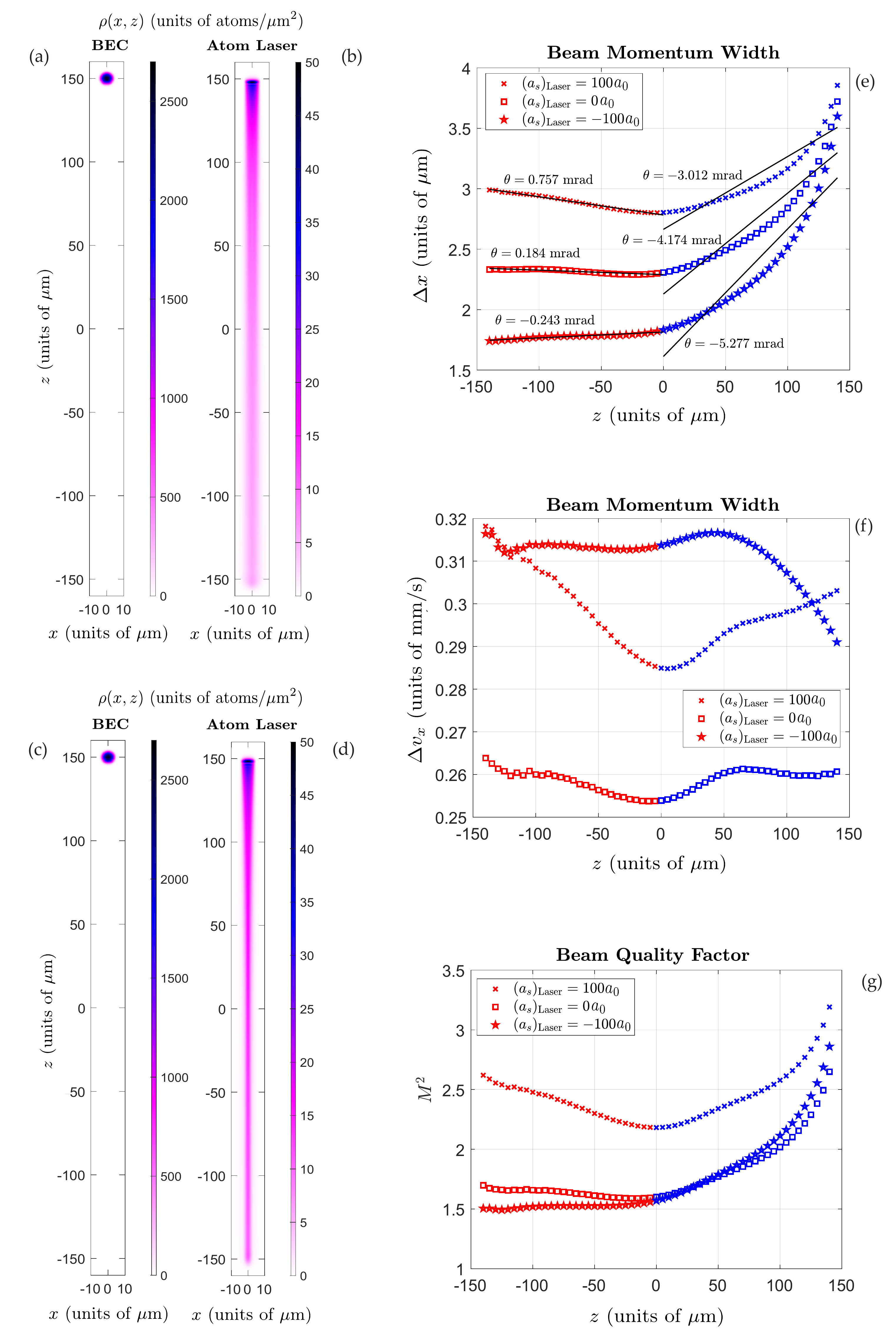}
		\caption{BEC [(a) and (c)] and atom laser [(b) and (d)] density profiles, indicated by the color map, in position space in the $x-z$ plane (cross section view), where $(a_s)_{\text{Laser}}=100a_0$ in (a) and (b), and $(a_s)_{\text{Laser}}=-100a_0$ in (c) and (d). The profile has been integrated over the $y$ axis in position space. The atom laser is outcoupled from the BEC using two Bragg photons, kicked by $p=2\hbar k$ and travels for a total distance of $d_z=300~\mu$m. Graphs (e), (f) and (g), respectively, show the results for beam width, beam momentum (transverse velocity) width and beam quality factor as a function of longitudinal path, $z$, for three different \textit{s}-wave interaction strengths, $(a_s)_{\text{Laser}}=100a_0$, $0$ and $-100a_0$, within the atom laser beam. In (e), (f) and (g), the Fresnel-Kirchhoff ($0\leq z\leq 140~\mu$m) and paraxial ($-140\leq z<0~\mu$m) zones are distinguished by the blue and red colors respectively. Each black solid line in graph (e) is a linear fit to the corresponding curve estimating the associated beam divergence, $\theta$, in both zones. Parameters used in the simulations are: $N_0=10^5$, $(a_s)_{\text{BEC}}=100a_0$,  $p=2\hbar k$, $\Delta_z=50$ nm, $\Omega=0.669$ kHz, $a_0=5.29\times 10^{-11}$ m and $K=4\times 10^{-41}$m$^6$s$^{-1}$.}
		\label{f513}
	\end{figure*}

	We consider a cigar-shaped $^{85}$Rb BEC elongated along the loose trap axis, $y$, while having a circular symmetry along the $x$ and $z$ axes following \cite{5_13, 5_30}. The trapping potential providing such a BEC geometry includes the tight radial frequencies of $\omega_x=\omega_z=2\pi\times 70$ Hz and axial frequency of $\omega_y=2\pi\times 10$ Hz. The BEC is set to have a fixed scattering length of $(a_s)_{\text{BEC}}=100a_0=5.291$ nm and is loaded with a constant flux such that the number of atoms, $N_0=10^5$, remains constant. To produce a quasi-continuous atom laser beam we choose two continuous Bragg pulses of the wavelength $\lambda=780.027$ nm propagating at an angle of $\alpha=180^{\circ}$ towards each other resulting in $|\mathbf{q}|=1.611\times 10^7$ m$^{-1}$. Secondly, we adjust the default outcoupling resonance width to be $\Delta_z$=50 nm requiring a constant Rabi frequency of $\Omega=0.669$ kHz acquired by $\Delta_z=-2g/\omega_z^2+2\sqrt{g^2/\omega_z^4+\hbar\Omega/m\omega_z^2}$ introduced in \cite{5_11}. The BEC and atom laser density profiles in the $(x, z)$ plane, $\rho(x, z)$, in position space within a two-photon Bragg process when the \textit{s}-wave scattering length within the beam is set to $(a_s)_{\text{Laser}}=100a_0$ and $(a_s)_{\text{Laser}}=-100a_0$ are displayed in Figs~\ref{f513} (a, b) and (c, d) respectively. The beam is Bragg-kicked by $p=\hbar|\mathbf{q}|=2\hbar k$, outcoupled from the BEC, and travels a total distance of $d_z=300~\mu$m, from $z_1=150~\mu$m (where the outcoupling resonance is located) to $z_2=-150~\mu$m [see Fig~\ref{f513} (b) and (d)]. In each simulation, the atom laser starts propagating at an initial velocity of $v_i=1.18$ cm/s while it completes the traveled path, at a final velocity of $v_f=7.76$ cm/s.

	For a free propagating atom laser, we have provided the associated results for the beam width, $\Delta x$, beam momentum width, $\Delta v_x$, and beam quality factor \cite{5_13}, $M^2=\frac{2}{\hbar}\Delta x\Delta p_x$ (where $\Delta p_x=m\Delta v_x$) for $(a_s)_{\text{Laser}}=100a_0, 0, -100a_0$ in Figs~\ref{f513} (e, f, g). As illustrated in the three graphs, the outcomes of the Fresnel-Kirchhoff and paraxial regions are distinguished by the blue ($0\leq z\leq 140~\mu$m) and red ($-140 \leq z<0~\mu$m) curves respectively. As expected, there is a striking difference in trend of the results between the two regions, which justifies the reason for classifying the beam evolution investigation into these individual regimes. Although the beam width for a repulsive, $(a_s)_{\text{Laser}}=100a_0$, a non-interacting, $(a_s)_{\text{Laser}}=0$, and an attractive, $(a_s)_{\text{Laser}}=-100a_0$, atom laser beam has a universal descending trend in the Fresnel-Kirchhoff regime. It undergoes a change in the paraxial zone depending on the beam interaction strength. For smaller and negative values of \textit{s}-wave interactions (i.e.  $(a_s)_{\text{Laser}}=0a_0$ or $-100a_0$), the beam width declines more significantly by traveling distance in the Kirchhoff regime (blue curves) compared to larger and positive interactions (i.e.  $(a_s)_{\text{Laser}}=100a_0$) [see Fig~\ref{f513} (e)]. The solid black lines provide a linear fit to the curves measuring the beam divergence angle for each specific interaction strength such that the sharpest angle, $\theta=-5.277$ mrad, relates to the most attractive atom laser beam, $(a_s)_{\text{Laser}}=-100a_0$, and the lowest angle, $\theta=-3.012$ mrad corresponds to the repulsive beam, $(a_s)_{\text{Laser}}=100a_0$. However, the situation is noticeably different in the paraxial regime. When setting the beam scattering length to $(a_s)_{\text{Laser}}=100a_0$, the beam starts diverging once it enters the paraxial region yielding a beam divergence angle of $\theta=0.757$ mrad. While this angle decreases to $\theta=0.184$ mrad by turning down the beam scattering length to $(a_s)_{\text{Laser}}=0a_0$, it is still positive meaning that the beam is slightly diverging in the paraxial regime. Nevertheless, it appears that negative interactions provide a converging beam as the angular divergence of negative, $\theta=-0.243$ mrad, when $(a_s)_{\text{Laser}}=-100a_0$. The resultant beam widths at $z=-140~\mu$m within the paraxial regime are achieved as $\Delta x= 2.990, 2.332$ and $1.741~\mu$m for $(a_s)_{\text{Laser}}=100a_0, 0$ and $-100a_0$ respectively.
	
			\begin{figure}
		\hskip -10ex
		\includegraphics[width=9cm, height=12cm,angle=0] {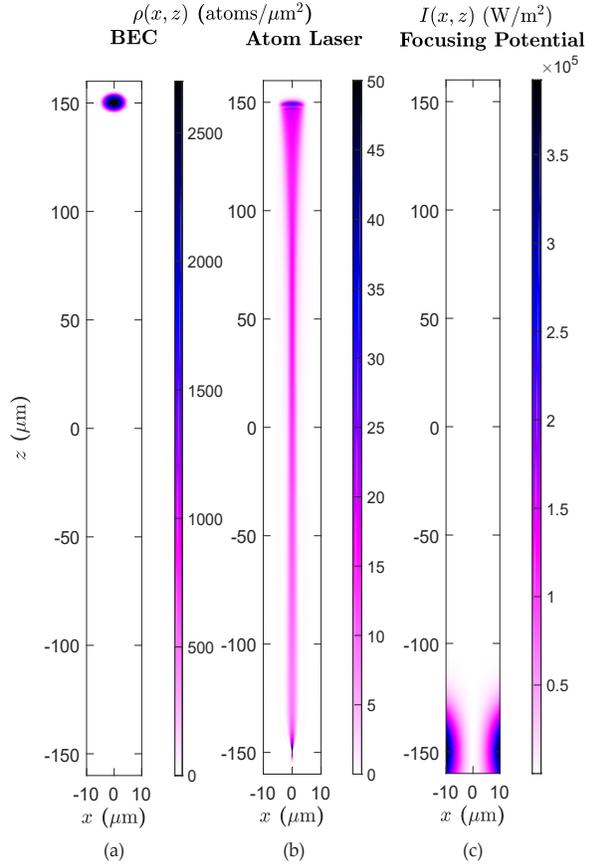}
		\caption{Simulation of an atom laser beam outcoupled from a BEC optimally focused by a harmonic optical focusing potential. (a, b): The trapped BEC and atom laser density profiles highlighted be the corresponding color maps. (c): The focusing potential intensity profile determined by the color map. The BEC and atom laser profiles have been integrated over the $y$ axis. The atom laser is outcoupled from the BEC at $z=150~\mu$m by a Bragg kick of $p=2\hbar k$ corresponding to an initial velocity of $v_i=1.18$ cm/s. It is then accelerated under gravity and arrives by $v_f=7.76$ cm/s at $z=-150~\mu$m to be focused by the potential whose maximum intensity is $I_0=9.91\times 10^6$ W/m$^2$ at $x=\pm \lambda/4=\pm 78~\mu$m. The parameters used in the simulation are: $N_0=10^5$, $\Delta_z=40$ nm, $\Omega=0.535$ kHz, $\sigma_z=25~\mu$m, $\lambda=312~\mu$m, $I_s=16.7$ W/m$^2$, $\gamma=38$ MHz, $\Delta=200$ GHz, $(a_s)_{\text{BEC}}=100a_0$, $(a_s)_{\text{Laser}}=-300a_0$,  $a_0=5.29\times 10^{-11}$ m and $K=4\times 10^{-41}$m$^6$s$^{-1}$.}
		\label{f521}
	\end{figure}

			\begin{figure*}
	\centering
	\includegraphics[width=15cm, height=15cm,angle=0] {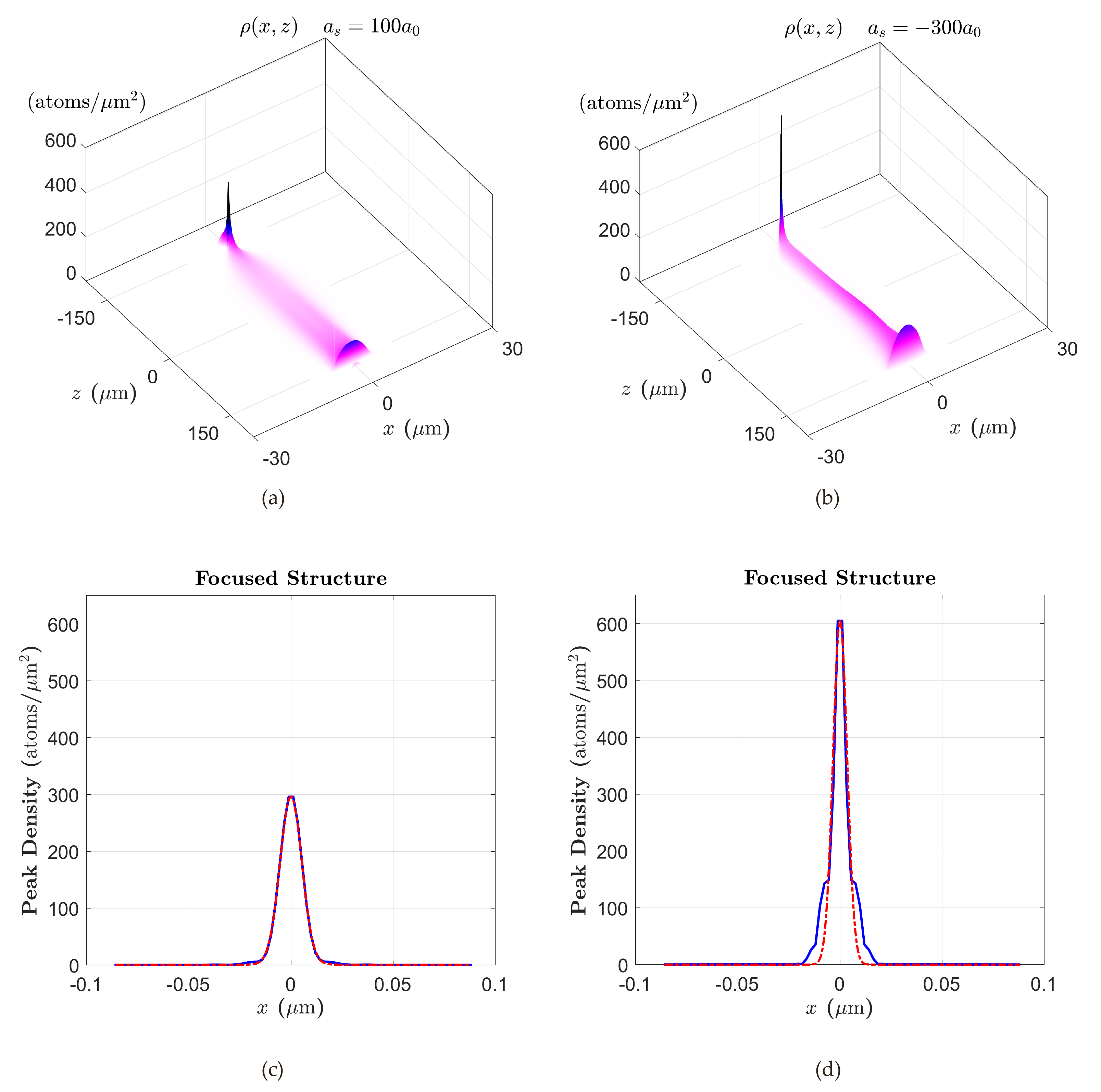}
	\caption{3D view of (a) $[a_s]_{\text{Laser}}=100a_0$, and (b) $[a_s]_{\text{Laser}}=-300a_0$ atom laser profiles in the focusing process as well as their focused profiles along the $x$ axis at $z=-150~\mu$m for (c) $[a_s]_{\text{Laser}}=100a_0$, and (d) $[a_s]_{\text{Laser}}=-300a_0$. In graphs (c) and (d), the solid blue curves indicate the 1D density profiles and the dashed red curves illustrate the Gaussian fits. The values of peak density and FWHM for $[a_s]_{\text{Laser}}=100a_0$ and $[a_s]_{\text{Laser}}=-300a_0$ are respectively achieved as $296.3$ atoms/$\mu$m$^2$, $13.2$ nm and $605.1$ atoms/$\mu$m$^2$, $8.04$ nm. The parameters used in the simulations are: $N_0=10^5$, $\Delta_z=40$ nm, $\Omega=0.535$ kHz, $\sigma_z=25~\mu$m, $\lambda=312~\mu$m, $I_s=16.7$ W/m$^2$, $\gamma=38$ MHz, $\Delta=200$ GHz, $(a_s)_{\text{BEC}}=100a_0$, $(a_s)_{\text{Laser}}=-300a_0$, $a_0=5.29\times 10^{-11}$ m and $K=4\times 10^{-41}$m$^6$s$^{-1}$.}
	\label{f522}
\end{figure*}

	We now consider the beam momentum width for the same interaction strengths. Figure~\ref{f513} (f) illustrates the variation of transverse velocity width, $\Delta v_x$, against the longitudinal path, $z$. Overall, one can clearly understand that $\Delta v_x$ for the case of a non-interacting beam, $(a_s)_{\text{Laser}}=0$, has the smallest values over the traveling path ($\Delta v_x=0.260$ and $0.263$ mm/s at  $z=140$ and $-140~\mu$m respectively). However, boosting the interaction strength, whether positively or negatively, results in an alteration in transverse velocities so that for stronger interactions (i.e. $(a_s)_{\text{Laser}}=100a_0$ and $-100a_0$), the variation of $\Delta v_x$ becomes significant. In the Kirchhoff zone, the beam transverse velocity width has an upward and downward trend by distance for $(a_s)_{\text{Laser}}=-100a_0$ and $(a_s)_{\text{Laser}}=100a_0$ respectively leading to greater values for an attractive atom laser beam. This implies that larger changes in transverse velocities occur for an attractive beam than for a repulsive one over the Kirchhoff region. In the paraxial regime, while $\Delta v_x$ has an ascending trend for $(a_s)_{\text{Laser}}=100a_0$, it slightly decreases for an attractive beam so that they almost tend to the same value, $\Delta v_x\sim 0.317$ mm/s, at $z=-140~\mu$m.
	
	To study the beam spatial mode, the results for the beam quality factor, $M^2$, as a function of longitudinal distance are shown in Fig~\ref{f513} (g) using the associated data from $\Delta x$ and $\Delta p_x$ [see Figs~\ref{f513} (e) and (f)]. According to the graph, in the paraxial zone, the largest and smallest values for the beam quality parameter are acquired when the interactions are set to $(a_s)_{\text{Laser}}=100a_0$ and $(a_s)_{\text{Laser}}=-100a_0$ respectively meaning that an attractive beam can provide one with a better beam spatial mode since the related $M^2$ values are closer to the minimum Heisenberg uncertainty. While the trend is almost the case at late stages of the Kirchhoff zone (i.e. $0\leq z\leq 50~\mu$m), it is seen that the $M^2$ results for $(a_s)_{\text{Laser}}=-100a_0$ are slightly greater than those of $(a_s)_{\text{Laser}}=0$ in a distance interval of $50< z\leq 140~\mu$m. Last but not least, similar to the beam widths in the paraxial regime, for $(a_s)_{\text{Laser}}=100a_0$, $0$ and $-100a_0$, respectively, a considerable increase (leading to $M^2=2.62$ at $z=-140~\mu$m), a slight increase (leading to $M^2=1.69$ at $z=-140~\mu$m) and a smooth descent (leading to $M^2=1.5$ at $z=-140~\mu$m) is observed in the beam quality outputs. However, they all decrease in the Kirchhoff region signifying that the beam quality improves as it travels further from the bottom of the condensate within this zone.

\subsection{Atom Laser Properties with Focusing}

	We now consider the presence of the focusing potential in estimating the dynamics of a propagating atom laser. Figs~\ref{f521}(a, b ,c) depict an atom laser that is Bragg kicked by $p=\hbar|\mathbf{q}|=2\hbar k$ outcoupled at $z=150~\mu$m from the outcoupling surface of a thickness of $\Delta_z=40$ nm, and is aimed to be optimally focused at $z=-150~\mu$m [see Figs~\ref{f521}(a, b), the BEC and atom laser density profiles]. The focusing potential field is red-detuned by $\Delta=200$ GHz from the $5~^2P_{3/2}$ level in the $^{85}$Rb D$_2$ transition while it consists of a wavelength of $\lambda=400\lambda_{\text{D}_2}=312~\mu$m and a radius size of $\sigma_z=25~\mu$m. Given the values of $I_s=16.7$ W/m$^2$ and $\gamma=38$ MHz for the $^{85}$Rb D$_2$ line, selecting $\xi=5.37$ provides an optimal power of $P_{2\hbar k}=2.433$ mW for focusing the beam at the center of potential, $z=-150~\mu$m [see Eq.(\ref{e29})]. The maximum intensity value at the two adjacent peaks separated by $\lambda/2=156~\mu$m is estimated as $I_0=8P_{2\hbar k}/\pi\sigma_z^2=9.91\times 10^6$ W/m$^2$  [see Fig~\ref{f521}(c), the focusing potential intensity profile]. In this example, the scattering length within the BEC and atom laser is chosen as $(a_s)_{\text{BEC}}=100a_0$ and $(a_s)_{\text{Laser}}=-300a_0$ such that the latter provides a self-focusing for the beam in addition to the focusing arising from the external optical potential.

	To have a distinct understanding of the self-focusing effect, we have also carried out simulations for a repulsive beam, $(a_s)_{\text{Laser}}=100a_0$ using the same parameters. Figs~\ref{f522}(a, b) show a 3D view of the atom laser density profile focused optimally at $z=-150~\mu$m when setting $(a_s)_{\text{Laser}}=100a_0$ and $-300a_0$ respectively. Applying an appropriate Gaussian fit to each focused profile at the focal point ($z=-150~\mu$m) along the $x$ axis, one can estimate the value of Full Width at Half Maximum (FWHM) for both repulsive [see Fig~\ref{f522}(c)] and attractive [see Fig~\ref{f522}(d)] beams. This results in  $13.2$ nm and $8.04$ nm linewidths respectively, which indicates that an attractive beam provides a superior resolution. Furthermore, as seen through Figs~\ref{f522}(a, b, c, d), focusing a beam with a high value of negative scattering length results in a significantly higher peak density, which for our case is obtained as $605.1$ atoms/$\mu$m$^2$ for $(a_s)_{\text{Laser}}=-300a_0$ compared to $296.3$ atoms/$\mu$m$^2$ for $(a_s)_{\text{Laser}}=100a_0$.

		\subsubsection{Impact of Varying the Focusing Potential parameters}
	
	We now investigate the influence of using various focusing potential parameters such as different potential radii and wavelengths. Since the potential slit size is estimated by $D=\lambda/2$, varying such a factor enhances focusing. Moreover, factors such as the potential power and maximum intensity are dependent on the wavenumber and radius of the potential. Hence, evaluating these possibilities would deliver a clear prospective of an optimal focusing potential.
	
		\begin{figure}
		\hskip -5ex
		\includegraphics[width=10cm, height=13cm,angle=0] {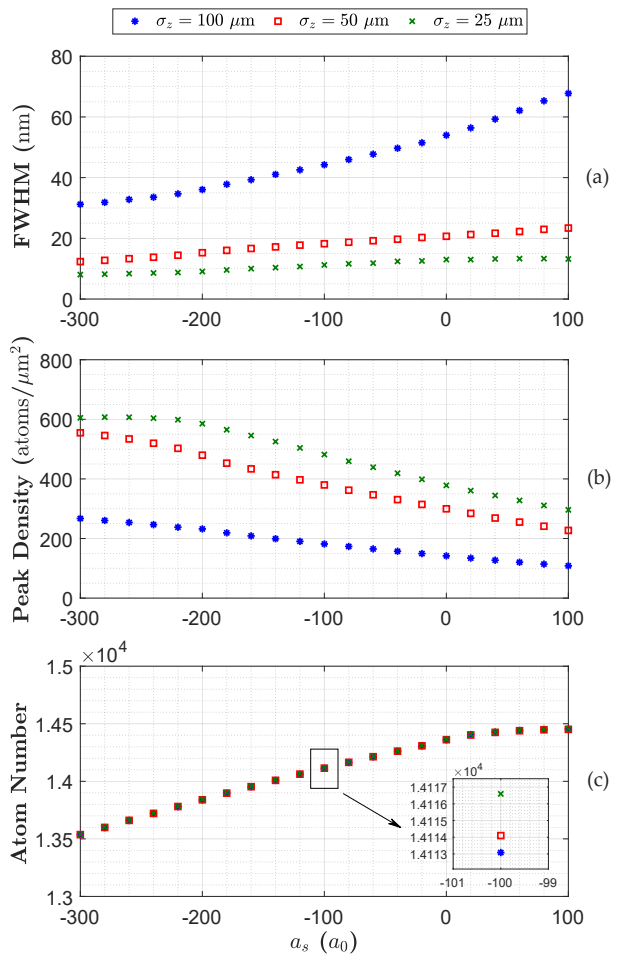}
		\caption{Results for (a) FWHM, (b) peak density, and (c) atom number within the focused atom laser beam as a function of the atom laser interaction strength. The blue dots, red squares and green crosses refer to $\sigma_z=100$, $50$ and $25~\mu$m respectively. All data for FWHM and peak density magnitudes are taken at the optimal focus spot, $z=-150~\mu$m, and the values for beam atom numbers are obtained by integrating the square absolute value of density profile from $z=150$ to $-150~\mu$m. The simulation parameters are: $N_0=10^5$, $\Delta_z=40$ nm, $\Omega=0.535$ kHz, $p=2\hbar k$, $\lambda=312~\mu$m, $I_s=16.7$ W/m$^2$, $\gamma=38$ MHz, $\Delta=200$ GHz, $(a_s)_{\text{BEC}}=100a_0$, $a_0=5.29\times 10^{-11}$ m and $K=4\times 10^{-41}$m$^6$s$^{-1}$.}
		\label{f523}
	\end{figure}	
	
	\subsubsection*{1.1. Potential Radius Size}

	For this purpose, we first conduct three simulations associated with different potential radii chosen as $\sigma_z=100$, $50$ and $25~\mu$m while keeping the slit size, $D=156~\mu$m, fixed. The \textit{s}-wave interactions within the atom laser beam are varied over a large range between  $(a_s)_{\text{Laser}}=100a_0$ and $=-300a_0$. Other parameters concerned with the BEC, Bragg outcoupling process and focusing potential are considered as constant values: $(a_s)_{\text{BEC}}=100a_0$; $\Delta_z=40$ nm; $\Omega=0.535$ kHz; $p=2\hbar k$ and $\Delta=200$ GHz. In spite of the fact that the optimal power required to focus the beam at $z=-150~\mu$m is independent of the potential radius size calculated as $P_{2\hbar k}=2.433$ mW, the maximum potential intensity takes a unique value for each radius size and is estimated as $I_s=6.19\times 10^5$, $2.47\times 10^6$ and $9.91\times 10^6$ W/m$^2$ for $\sigma_z=100$, $50$ and $25~\mu$m respectively.
	
	Figs~\ref{f523}(a, b, c) depict the outcomes for the beam FWHM, peak density and atom number. At a glance, one can notice that for any atom laser interaction strength, the leading resolutions (lower FWHMs) and higher peak densities, indicated by green crosses, are obtained using smaller potential radii such that the best case scenario occurs for $\sigma_z=25~\mu$m. This is because lower radius sizes result in relatively higher potential field intensities enhancing the dipole force. In addition, for every radius size, decreasing the magnitude of interaction from positive to negative values leads to a better resolution and higher peak density due to the self-focusing effect. For instance, while utilizing $\sigma_z=100$, $50$ and $25~\mu$m bring $67.76$, $23.38$ and $13.2$ nm linewidths for $(a_s)_{\text{Laser}}=100a_0$, these radii would yield $31.21$, $12.33$ and $8.04$ nm FWHMs for $(a_s)_{\text{Laser}}=-300a_0$. Turning to the peak density, one can observe that the values for a focused atom laser of $(a_s)_{\text{Laser}}=100a_0$, can reach up to $108$, $229$ and $296$ atoms/$\mu$m$^2$ for $\sigma_z=100$, $50$ and $25~\mu$m respectively, while the peak densities for a focused beam of $(a_s)_{\text{Laser}}=-300a_0$ are acquired as $267.1$, $554.9$ and $605.1$ atoms/$\mu$m$^2$ respectively.
	
	In studying the results for the number of atoms within the beam, one can understand that reducing the scattering length causes relatively lower atoms to remain in the system during the focusing event since three-body losses become considerable for attractive atom laser beams in the high density regimes (i.e. the focusing regimes). However, for a certain interaction strength, altering the potential radius does not considerably impact the total beam atom number. As a case in point, setting the beam \textit{s}-wave interaction to $(a_s)_{\text{Laser}}=-100a_0$ brings $14113$, $14114$ and $14117$ atoms within the focused beam for $\sigma_z=100$, $50$ and $25~\mu$m respectively.
	
	\begin{figure}[t!]
		\hskip -5ex
		\includegraphics[width=10cm, height=13cm,angle=0] {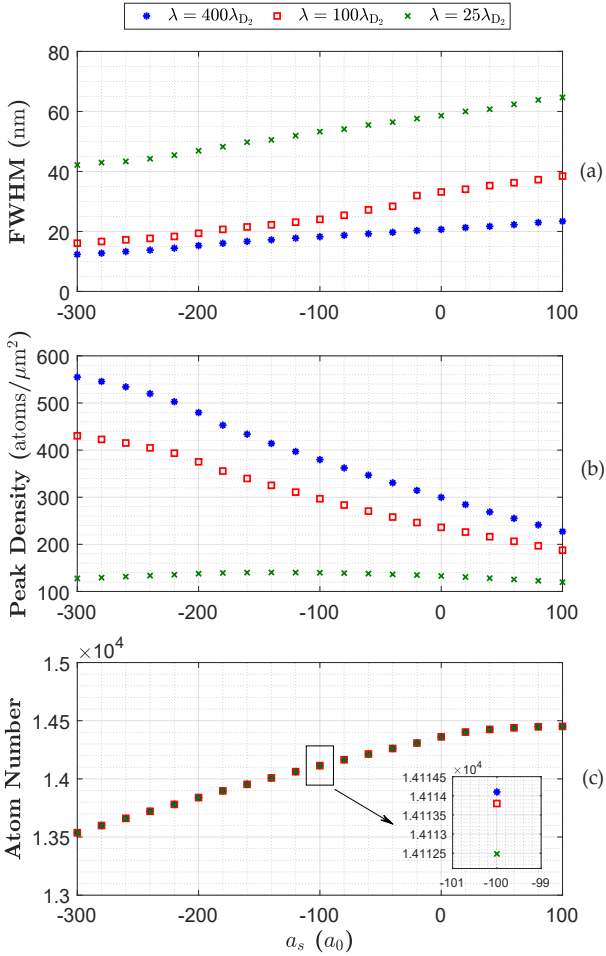}
		\caption{Results for (a) FWHM, (b) peak density, and (c) atom number within the focused atom laser beam as a function of the atom laser interaction strength. The blue dots, red squares and green crosses refer to $\lambda=400\lambda_{\text{D}_2}$, $100\lambda_{\text{D}_2}$ and $25\lambda_{\text{D}_2}$ corresponding to $D=156$, $39$ and $19.5~\mu$m respectively. All data for FWHM and peak density magnitudes are taken at the optimal focus spot, $z=-150~\mu$m, and the values for beam atom numbers are obtained by integrating the square absolute value of density profile from $z=150$ to $-150~\mu$m. The simulation parameters are: $N_0=10^5$, $\Delta_z=40$ nm, $\Omega=0.535$ kHz, $p=2\hbar k$, $\sigma_z=50~\mu$m, $I_s=16.7$ W/m$^2$, $\gamma=38$ MHz, $\Delta=200$ GHz, $(a_s)_{\text{BEC}}=100a_0$, $a_0=5.29\times 10^{-11}$ m and $K=4\times 10^{-41}$m$^6$s$^{-1}$.}
		\label{f524}
	\end{figure}

	\subsubsection*{1.2. Potential slit Size}

	For the second category of simulations, the focusing potential takes a fixed radius value of $\sigma_z=50~\mu$m whereas its slit size varies over $400\lambda_{\text{D}_2}/2$, $100\lambda_{\text{D}_2}/2$ and $25\lambda_{\text{D}_2}/2$, which are equal to $D=156$, $39$ and  $9.75~\mu$m.
	Again, applying the same parameters for the BEC, Bragg scattering process and laser detuning as used for the previous case, the resultant optimal powers for focusing at $z=-150~\mu$m given the slit sizes $D=156$, $39$ and  $9.75~\mu$m are, respectively, estimated as $P_{2\hbar k}=2.433$ mW, $152~\mu$W and $9.5~\mu$W corresponding to the maximum Gaussian intensity $I_0=2.47\times 10^6$, $1.59\times 10^5$ and $9.68\times 10^3$ W/m$^2$.
	
	The related outcomes for FWHM, peak density and atom number of the focused beam are illustrated in Figs~\ref{f524}(a, b, c). As seen, for each atom laser scattering length, the most broad structure linewidth and the lowest peak density (indicated by the blue dot points) are caused by the narrowest potential slit, $D=19.5~\mu$m ($\lambda=25\lambda_{\text{D}_2}$) since this value is associated with the lowest potential power and intensity compared to the other two slit sizes resulting in a poorer focus. It is clear that enlarging the slit size requires an increase in power and consequently in intensity to optimally focus the beam to the same spot. Furthermore, making the beam more attractive is the second factor to achieve a better focusing since exerting relatively low or negative scattering lengths improves the resolution and peak density for each slit size. For example, a gap of $D=156~\mu$m ($\lambda=400\lambda_{\text{D}_2}$) leads to $12.33$ nm and $554.9$ atoms/$\mu$m$^2$ in FWHM and peak density for $(a_s)_{\text{Laser}}=-300a_0$ whereas these values are obtained as $23.38$ nm and $227.1$ atoms/$\mu$m$^2$ for $(a_s)_{\text{Laser}}=100a_0$.
	
	However, the impact of the slit size is almost negligible on the number of atoms within the beam. This is similar to the previous case where changing the potential radius size would not make much difference to atom numbers for a certain value of $(a_s)_{\text{Laser}}$. Hence, one could infer that the beam atom number is independent of the potential power and intensity value, and it is only determined by the interaction strength such that highly negative scattering lengths would yield lower atoms in the system due to higher loss rate.

	\begin{figure}
		\hskip -5ex
		\includegraphics[width=10cm, height=13cm,angle=0] {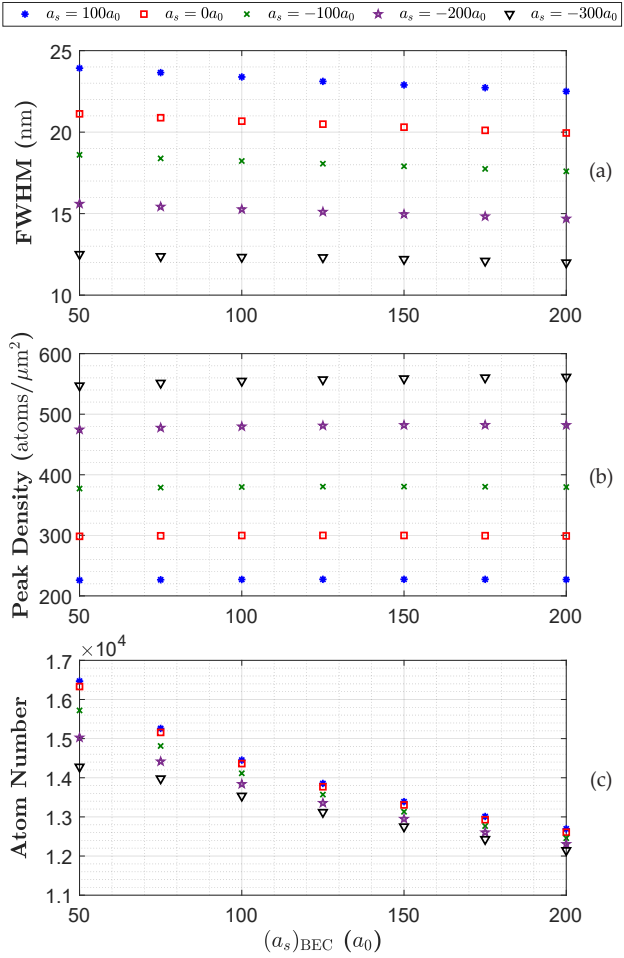}
		\caption{Results for (a) FWHM, (b) peak density, and (c) atom number within the focused atom laser beam as a function of the BEC scattering length. The blue dots, red squares, green crosses, purple stars and black triangles correspond to $(a_s)_{\text{Laser}}=100a_0$, $0$, $-100a_0$, $-200a_0$ and $-300a_0$ respectively. All data for FWHM and peak density magnitudes are evaluated at the optimal focus spot, $z=-150~\mu$m, and the values for beam atom numbers are obtained by integrating the square absolute value of density profile from $z=150$ to $-150~\mu$m. The simulation parameters are: $N_0=10^5$, $\Delta_z=40$ nm, $\Omega=0.535$ kHz, $p=2\hbar k$, $\sigma_z=50~\mu$m, $\lambda=400\lambda_{\text{D}_2}$, $I_s=16.7$ W/m$^2$, $\gamma=38$ MHz, $\Delta=200$ GHz, $a_0=5.29\times 10^{-11}$ m and $K=4\times 10^{-41}$m$^6$s$^{-1}$.}
		\label{f525}
	\end{figure}

		\subsubsection{Impact of Varying the BEC and Bragg Scattering Parameters}
	
	It is important to consider the influence of using a range of different outcoupling strengths and momentum kicks in the Bragg scattering event as well as various interaction strengths within the BEC on the dynamics of a focused atom laser. In this section, we keep the geometrical parameters constant for the focusing potential, $\sigma_z=50~\mu$m and $D=156~\mu$m, while the factors with regard to the BEC and atom laser are varied.\\
	
		\subsubsection*{2.1. BEC Scattering Length}
		
		We tune the BEC interaction strength over $50a_0\leq (a_s)_{\text{BEC}}\leq200a_0$ such that for every $(a_s)_{\text{BEC}}$, five different atom laser scattering lengths, $(a_s)_{\text{Laser}}=100a_0$, $0$, $-100a_0$, $-200a_0$ and $-300a_0$ are considered. The Bragg outcoupling parameters are kept constant: $\Delta_z=40$ nm; $\Omega=0.535$ kHz; $p=2\hbar k$. This necessitates the focusing potential power and maximum intensity to be $P_{2\hbar k}= 2.433$ mW and $I_0= 2.47\times 10^6$ W/m$^2$ to optimally bring the beam at the focal spot, $z=-150~\mu$m.
	
	The results of our numerical simulations for the magnitudes of beam FWHM, peak density and atom number are represented in Figs~\ref{f525}(a, b, c). Firstly, one can observe that the change of BEC interaction strength does not significantly influence the focal spot sizes as well as the peak densities. For each atom laser scattering length, there is a very slight reduction and rise, respectively, in FWHM and peak density from $(a_s)_{\text{BEC}}=50a_0$ to $200a_0$ meaning that larger $(a_s)_{\text{BEC}}$ can negligibly boost the resolution and peak density. Secondly, as expected, lowering the beam interaction strength results in enhanced resolutions and peak densities for any $(a_s)_{\text{BEC}}$. To illustrate, considering $(a_s)_{\text{Laser}}=-300a_0$, the resultant linewidths for $(a_s)_{\text{BEC}}=50a_0$ and $200a_0$ are, respectively, as $12.51$ nm and $11.99$ nm while the focused peaks are scaled as $547.1$ and $561.5$ atoms/$\mu$m$^2$.
	
	Nevertheless, various BEC \textit{s}-wave interactions influence the number of atoms within the beam such that turning up $(a_s)_{\text{Laser}}$ yields less atoms within the beam during the focusing event implying that more recombination losses occur. That is to say for $(a_s)_{\text{Laser}}=100a_0$, the beam atom number diminishes from $16469$ to $12694$ when switching the $(a_s)_{\text{BEC}}$ value from $50a_0$ to $200a_0$. Lastly, the change in atom number values between different beam interaction strengths for a certain $(a_s)_{\text{BEC}}$ results from the three-body losses such that the more attractive beam, the more atom loss. The change, however, reduces by expanding the BEC scattering length. All in all, we can conclude that lowering $(a_s)_{\text{BEC}}$ could be advantageous only to the number of atoms whereas the focused profile is almost independent of a choice of BEC interaction strength.
	
		\begin{figure}
		\hskip -5ex
		\includegraphics[width=10cm, height=13cm,angle=0] {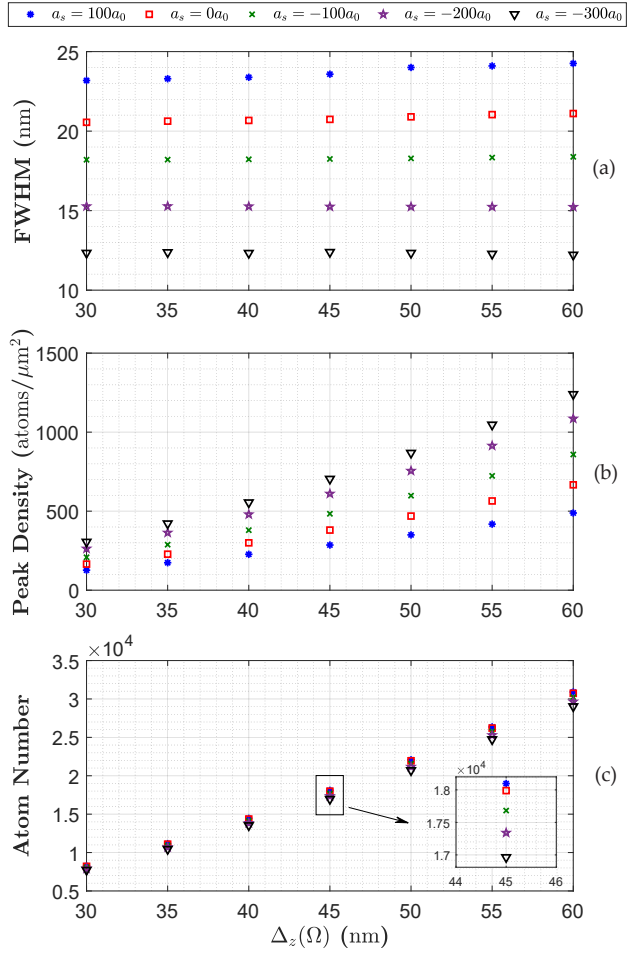}
		\caption{Results for (a) FWHM, (b) peak density, and (c) atom number within the focused atom laser beam as a function of the outcoupling resonance. The blue dots, red squares, green crosses, purple stars and black triangles correspond to $(a_s)_{\text{Laser}}=100a_0$, $0$, $-100a_0$, $-200a_0$ and $-300a_0$ respectively. All data for FWHM and peak density magnitudes are evaluated at the optimal focus spot, $z=-150~\mu$m, and the values for beam atom numbers are obtained by integrating the square absolute value of density profile from $z=150$ to $-150~\mu$m. The simulation parameters are: $N_0=10^5$, $(a_s)_{\text{BEC}}=100a_0$, $\Omega=0.535$ kHz, $p=2\hbar k$, $\sigma_z=50~\mu$m, $\lambda=400\lambda_{\text{D}_2}$, $I_s=16.7$ W/m$^2$, $\gamma=38$ MHz, $\Delta=200$ GHz, $a_0=5.29\times 10^{-11}$ m and $K=4\times 10^{-41}$m$^6$s$^{-1}$.}
		\label{f526}
	\end{figure}
	
	\begin{figure}
		\hskip -5ex
		\includegraphics[width=10cm, height=13cm,angle=0] {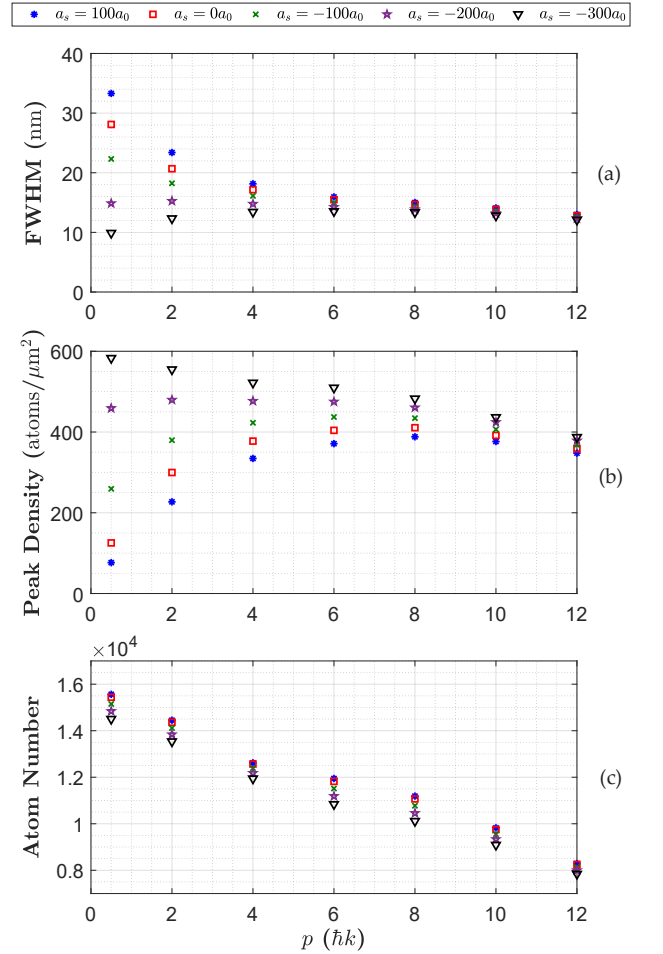}
		\caption{Results for (a) FWHM, (b) peak density, and (c) atom number within the focused atom laser beam as a function of the momentum kick. The blue dots, red squares, green crosses, purple stars and black triangles correspond to $(a_s)_{\text{Laser}}=100a_0$, $0$, $-100a_0$, $-200a_0$ and $-300a_0$ respectively. All data for FWHM and peak density magnitudes are taken at the optimal focus spot, $z=-150~\mu$m, and the values for beam atom numbers are obtained by integrating the square absolute value of density profile from $z=150$ to $-150~\mu$m. The simulation parameters are: $N_0=10^5$, $(a_s)_{\text{BEC}}=100a_0$, $\Delta_z=40$ nm, $\Omega=0.535$ kHz, $\sigma_z=50~\mu$m, $\lambda=400\lambda_{\text{D}_2}$, $I_s=16.7$ W/m$^2$, $\gamma=38$ MHz, $\Delta=200$ GHz, $a_0=5.29\times 10^{-11}$ m and $K=4\times 10^{-41}$m$^6$s$^{-1}$.}
		\label{f527}
	\end{figure}
	
	\subsubsection*{2.2. Bragg Outcoupling Factors}

	Here, the effect of employing a wide range of outcoupling strengths as well as momentum kicks is examined on a focused atom laser beam. For the first group of simulations, we suppose the outcoupling resonance value to be in the range $30\leq\Delta_z\leq 60$ nm corresponding to the two-photon Rabi frequency of $0.393\leq\Omega\leq 0.786$ kHz given that the tight trap frequency along the falling direction is taken as $\omega_z=2\pi\times 70$ Hz. The atom laser is outcoupled by a two-photon process being kicked by $p=2\hbar k$, and the parameters associated with the interaction strength between the atoms within the BEC and as well as the potential structural factors are assumed to be constant: $(a_s)_{\text{BEC}}=100a_0$; $\sigma_z=50~\mu$m and $D=156~\mu$m. Likewise, the optimal potential power and peak intensity are required to be $P_{2\hbar k}= 2.433$ mW and $I_0= 2.47\times 10^6$ W/m$^2$ for the best possible focus at $z=-150~\mu$m.
	
	The results for the focusing are displayed in Figs~\ref{f526}(a, b, c). Probing the FWHM outcomes, one can understand that the resolution has a minimal dependence on the outcoupling Rabi frequency. In other words, using any strength of outcoupling for a certain momentum kick results in almost the same resolution for a specific $(a_s)_{\text{Laser}}$. However, this is not the case for the peak density values. Increasing the outcoupling strength generates higher focused peak densities as more atoms are extracted from the BEC contributing within the atom laser beam.
	
	Analyzing the atom number graph, one can notice that the quantity of beam atoms as the outcoupling strength is increased. While for any particular $\Delta_z$, there exists a slight difference in atom numbers resulting from various $(a_s)_{\text{Laser}}$, these values undergo a significant variation from $\Delta_z=30$ to $60$ nm. As an exemplar, for $(a_s)_{\text{Laser}}=100a_0$, the beam atom number rises from $8261$ to $30904$.

	We now perform the second class of simulations in which the momentum kick exerted to the beam is varied between $p=0.5\hbar k$ and $12\hbar k$. This immediately requires the potential power to have values from $P_{0.5\hbar k}=2.38$ mW (corresponding to $I_0=2.42\times 10^6$ W/m$^2$) to $P_{12\hbar k}=4.48$ mW (corresponding to $I_0=4.57\times 10^6$ W/m$^2$) provided that all other parameters of the focusing potential are the same as the previous simulation. By setting the outcoupling strength to $\Omega=0.524$ kHz ($\Delta_z=40$ nm) and considering $(a_s)_{\text{BEC}}=100a_0$, the associated results are gathered in Figs~\ref{f527}(a, b, c). The first point observed in Figs~\ref{f527}(a, b) is that applying higher momentum kicks causes the FWHMs and peak densities to tend towards a steady state regardless of the magnitude of interaction strength within the beam. This means that the focused profile of a beam with a high enough longitudinal velocity becomes independent of the beam inter atom-atom interactions. Another point extracted from the graphs (a) and (b) is that while a highly attractive beam [i.e. $(a_s)_{\text{Laser}}=-300a_0$] dramatically improves the resolution and peak density for relatively small momentum kicks, the consequence of using a highly repulsive beam [i.e. $(a_s)_{\text{Laser}}=100a_0$] is destructive in the focused profile for the same case. As an instance, for $p=0.5\hbar k$, tuning the beam \textit{s}-wave scattering length to $(a_s)_{\text{Laser}}=100a_0$ and $-300a_0$ would respectively produce the FWHMs and peak densities of $33.3$ nm, $76.96$ atoms/$\mu$m$^2$ and $9.89$ nm, $583$ atoms/$\mu$m$^2$.
	
	Last but not least, since enhancing the beam initial momentum kick would couple a lower fraction of atoms out of the condensate for a distinct value of $\Omega$, one can see a substantial drop in the beam atom number in Fig~\ref{f527}(c) from $p=0.5\hbar k$ to $12\hbar k$ for each $(a_s)_{\text{Laser}}$. One example is an atom laser of $(a_s)_{\text{Laser}}=-300a_0$ where the beam atom number reduces from $14503$ to $7843$. Overall, one could realize that imparting a relatively low value of momentum kick to a greatly attractive beam could be beneficial in atom laser focusing process.

	\section{Conclusions}
	
	In this paper, we investigated the focusing dynamics of an evolving quasi-continuous atom laser beam. We concentrated our attention on the Bragg outcoupling method in which the BEC is initially confined in an optical dipole trap. We proposed the two-state model based on the GPE for the BEC and atom laser to consider the inter-atomic interactions, three-body recombination losses as well as the focusing potential parameters. We examined the parameters of an atom laser such as the beam width, beam momentum width, beam divergence and the beam quality factor in both the Kirchhoff and paraxial regimes in the absence of a focusing potential. Moreover, the impact of altering various factors within the focusing potential, BEC and Bragg scattering on a focused beam was studied such as using different potential radius and slit sizes, setting a range of BEC \textit{s}-wave scattering lengths, exerting a number of momentum kicks and applying diverse outcoupling strengths.
	
	As a first conclusion, using relatively low or negative beam scattering lengths brings more convergent and narrower beams along with better quality factors. We perceived that lowering the beam scattering length always results in a higher resolution and peak density since attractive beams self-focus. However, this action negatively affects the atom number within the beam due to the three-body recombination losses. It was then found that smaller radii (i.e. $\sigma_z=25~\mu$m) and larger slits (i.e. $D=156~\mu$m) of the focusing potential create narrower and higher focused structures since such potential values necessitate larger optimal powers and peak intensities providing more robust focusing, specifically we demonstrated structures with $8.04$ nm and $605.1$ atoms/$\mu$m$^2$ in resolution and peak density respectively. Following the investigation of a focused beam, we observed that the FWHM and peak density values are almost independent of the variation of BEC scattering length. In exploring outcoupling strengths, we concluded that while the resolution of focused profiles may not have a considerable dependence on the value of $\Omega$, their peak density is strongly affiliated to $\Omega$ such that utilizing larger values is more beneficial leading to higher structures. Eventually, we realized that imparting relatively lower momentum kicks (i.e. $p\leq 4\hbar k$) improves the resolution and peak density of attractive beams (i.e. $(a_s)_{\text{laser}}=-300a_0$). Nonetheless, larger momentum kicks (i.e. $p=12\hbar k$) may cause the FWHM and peak density values to tend to a steady state for any $(a_s)_{\text{Laser}}$.\\
	
	\section*{ACKNOWLEDGMENTS}
	
	R.R. is supported by the Australian National University International Research Scholarship award. The authors would like to thank Timothy Senden and Hans A. Bachor for useful discussions and feedback. 
	
	\bibliography{RRef_1,RRef_2,RRef_49,RRef_50,RRef_51,RRef_52,R3_1,RI_1,RI_2,RI_3,RI_4,RI_5,RI_6,RI_7,RI_8,RI_9,RI_10,RI_11,RI_12,RI_13,RI_14,RI_15,RI_16,RI_17,RI_18,RI_19,RI_20,RI_21,RI_22,RI_23,RI_24,RI_25,RI_26,RI_27,RI_28,RI_29,RI_30,RI_31,RI_32,R5_1,R5_32,R5_6,R5_7,R5_13,R5_9,R5_10,R5_11,R5_12,R5_18,AL1,AL2,AL3,AL4,R5_31,R5_8,R3_8,R3_9,Ref_15,Ref_16,Ref_17,Ref_10,Ref_11,Ref_12,Ref_13,Ref_20,Ref_29,R5_21,R5_25,R5_26,R5_28,R5_33,R5_29,Ref_34,Ref_108,Ref_112,Ref_38,Ref_39,Ref_37,Ref_19,R5_30,Ri1,Ri2,Ri3,Ri4}

\begin{thebibliography}{71}%
\makeatletter
\providecommand \@ifxundefined [1]{%
 \@ifx{#1\undefined}
}%
\providecommand \@ifnum [1]{%
 \ifnum #1\expandafter \@firstoftwo
 \else \expandafter \@secondoftwo
 \fi
}%
\providecommand \@ifx [1]{%
 \ifx #1\expandafter \@firstoftwo
 \else \expandafter \@secondoftwo
 \fi
}%
\providecommand \natexlab [1]{#1}%
\providecommand \enquote  [1]{``#1''}%
\providecommand \bibnamefont  [1]{#1}%
\providecommand \bibfnamefont [1]{#1}%
\providecommand \citenamefont [1]{#1}%
\providecommand \href@noop [0]{\@secondoftwo}%
\providecommand \href [0]{\begingroup \@sanitize@url \@href}%
\providecommand \@href[1]{\@@startlink{#1}\@@href}%
\providecommand \@@href[1]{\endgroup#1\@@endlink}%
\providecommand \@sanitize@url [0]{\catcode `\\12\catcode `\$12\catcode
  `\&12\catcode `\#12\catcode `\^12\catcode `\_12\catcode `\%12\relax}%
\providecommand \@@startlink[1]{}%
\providecommand \@@endlink[0]{}%
\providecommand \url  [0]{\begingroup\@sanitize@url \@url }%
\providecommand \@url [1]{\endgroup\@href {#1}{\urlprefix }}%
\providecommand \urlprefix  [0]{URL }%
\providecommand \Eprint [0]{\href }%
\providecommand \doibase [0]{https://doi.org/}%
\providecommand \selectlanguage [0]{\@gobble}%
\providecommand \bibinfo  [0]{\@secondoftwo}%
\providecommand \bibfield  [0]{\@secondoftwo}%
\providecommand \translation [1]{[#1]}%
\providecommand \BibitemOpen [0]{}%
\providecommand \bibitemStop [0]{}%
\providecommand \bibitemNoStop [0]{.\EOS\space}%
\providecommand \EOS [0]{\spacefactor3000\relax}%
\providecommand \BibitemShut  [1]{\csname bibitem#1\endcsname}%
\let\auto@bib@innerbib\@empty
\bibitem [{\citenamefont {Metcalf}\ and\ \citenamefont {Van~der
  Straten}(2007)}]{I1}%
  \BibitemOpen
  \bibfield  {author} {\bibinfo {author} {\bibfnamefont {H.~J.}\ \bibnamefont
  {Metcalf}}\ and\ \bibinfo {author} {\bibfnamefont {P.}~\bibnamefont {Van~der
  Straten}},\ }\bibfield  {title} {\bibinfo {title} {Laser cooling and trapping
  of neutral atoms},\ }\href@noop {} {\bibfield  {journal} {\bibinfo  {journal}
  {The Optics Encyclopedia: Basic Foundations and Practical Applications}\ }
  (\bibinfo {year} {2007})}\BibitemShut {NoStop}%
\bibitem [{\citenamefont {Adams}\ and\ \citenamefont {Riis}(1997)}]{I2}%
  \BibitemOpen
  \bibfield  {author} {\bibinfo {author} {\bibfnamefont {C.}~\bibnamefont
  {Adams}}\ and\ \bibinfo {author} {\bibfnamefont {E.}~\bibnamefont {Riis}},\
  }\bibfield  {title} {\bibinfo {title} {Laser cooling and trapping of neutral
  atoms},\ }\href@noop {} {\bibfield  {journal} {\bibinfo  {journal} {Progress
  in quantum electronics}\ }\textbf {\bibinfo {volume} {21}},\ \bibinfo {pages}
  {1} (\bibinfo {year} {1997})}\BibitemShut {NoStop}%
\bibitem [{\citenamefont {Phillips}(1992)}]{I3}%
  \BibitemOpen
  \bibfield  {author} {\bibinfo {author} {\bibfnamefont {W.~D.}\ \bibnamefont
  {Phillips}},\ }\href@noop {} {\emph {\bibinfo {title} {Laser cooling and
  trapping of neutral atoms}}},\ \bibinfo {type} {Tech. Rep.}\ (\bibinfo
  {institution} {NATIONAL INST OF STANDARDS AND TECHNOLOGY GAITHERSBURG MD},\
  \bibinfo {year} {1992})\BibitemShut {NoStop}%
\bibitem [{\citenamefont {Ashkin}(1997)}]{I4}%
  \BibitemOpen
  \bibfield  {author} {\bibinfo {author} {\bibfnamefont {A.}~\bibnamefont
  {Ashkin}},\ }\bibfield  {title} {\bibinfo {title} {Optical trapping and
  manipulation of neutral particles using lasers},\ }\href@noop {} {\bibfield
  {journal} {\bibinfo  {journal} {Proceedings of the National Academy of
  Sciences}\ }\textbf {\bibinfo {volume} {94}},\ \bibinfo {pages} {4853}
  (\bibinfo {year} {1997})}\BibitemShut {NoStop}%
\bibitem [{\citenamefont {Weiss}\ and\ \citenamefont {Saffman}(2017)}]{I11}%
  \BibitemOpen
  \bibfield  {author} {\bibinfo {author} {\bibfnamefont {D.~S.}\ \bibnamefont
  {Weiss}}\ and\ \bibinfo {author} {\bibfnamefont {M.}~\bibnamefont
  {Saffman}},\ }\bibfield  {title} {\bibinfo {title} {Quantum computing with
  neutral atoms},\ }\href@noop {} {\bibfield  {journal} {\bibinfo  {journal}
  {Physics Today}\ }\textbf {\bibinfo {volume} {70}} (\bibinfo {year}
  {2017})}\BibitemShut {NoStop}%
\bibitem [{\citenamefont {Deutsch}\ \emph {et~al.}(2000)\citenamefont
  {Deutsch}, \citenamefont {Brennen},\ and\ \citenamefont {Jessen}}]{I12}%
  \BibitemOpen
  \bibfield  {author} {\bibinfo {author} {\bibfnamefont {I.~H.}\ \bibnamefont
  {Deutsch}}, \bibinfo {author} {\bibfnamefont {G.~K.}\ \bibnamefont
  {Brennen}},\ and\ \bibinfo {author} {\bibfnamefont {P.~S.}\ \bibnamefont
  {Jessen}},\ }\bibfield  {title} {\bibinfo {title} {Quantum computing with
  neutral atoms in an optical lattice},\ }\href@noop {} {\bibfield  {journal}
  {\bibinfo  {journal} {Fortschritte der Physik: Progress of Physics}\ }\textbf
  {\bibinfo {volume} {48}},\ \bibinfo {pages} {925} (\bibinfo {year}
  {2000})}\BibitemShut {NoStop}%
\bibitem [{\citenamefont {Horodecki}\ \emph {et~al.}(2009)\citenamefont
  {Horodecki}, \citenamefont {Horodecki}, \citenamefont {Horodecki},\ and\
  \citenamefont {Horodecki}}]{I14}%
  \BibitemOpen
  \bibfield  {author} {\bibinfo {author} {\bibfnamefont {R.}~\bibnamefont
  {Horodecki}}, \bibinfo {author} {\bibfnamefont {P.}~\bibnamefont
  {Horodecki}}, \bibinfo {author} {\bibfnamefont {M.}~\bibnamefont
  {Horodecki}},\ and\ \bibinfo {author} {\bibfnamefont {K.}~\bibnamefont
  {Horodecki}},\ }\bibfield  {title} {\bibinfo {title} {Quantum entanglement},\
  }\href@noop {} {\bibfield  {journal} {\bibinfo  {journal} {Reviews of modern
  physics}\ }\textbf {\bibinfo {volume} {81}},\ \bibinfo {pages} {865}
  (\bibinfo {year} {2009})}\BibitemShut {NoStop}%
\bibitem [{\citenamefont {Brazhnyi}\ and\ \citenamefont {Konotop}(2004)}]{I15}%
  \BibitemOpen
  \bibfield  {author} {\bibinfo {author} {\bibfnamefont {V.}~\bibnamefont
  {Brazhnyi}}\ and\ \bibinfo {author} {\bibfnamefont {V.}~\bibnamefont
  {Konotop}},\ }\bibfield  {title} {\bibinfo {title} {Theory of nonlinear
  matter waves in optical lattices},\ }\href@noop {} {\bibfield  {journal}
  {\bibinfo  {journal} {Modern Physics Letters B}\ }\textbf {\bibinfo {volume}
  {18}},\ \bibinfo {pages} {627} (\bibinfo {year} {2004})}\BibitemShut
  {NoStop}%
\bibitem [{\citenamefont {Scheppach}\ \emph {et~al.}(2010)\citenamefont
  {Scheppach}, \citenamefont {Berges},\ and\ \citenamefont {Gasenzer}}]{I16}%
  \BibitemOpen
  \bibfield  {author} {\bibinfo {author} {\bibfnamefont {C.}~\bibnamefont
  {Scheppach}}, \bibinfo {author} {\bibfnamefont {J.}~\bibnamefont {Berges}},\
  and\ \bibinfo {author} {\bibfnamefont {T.}~\bibnamefont {Gasenzer}},\
  }\bibfield  {title} {\bibinfo {title} {Matter-wave turbulence: Beyond kinetic
  scaling},\ }\href@noop {} {\bibfield  {journal} {\bibinfo  {journal}
  {Physical Review A}\ }\textbf {\bibinfo {volume} {81}},\ \bibinfo {pages}
  {033611} (\bibinfo {year} {2010})}\BibitemShut {NoStop}%
\bibitem [{\citenamefont {Canuel}\ \emph {et~al.}(2006)\citenamefont {Canuel},
  \citenamefont {Leduc}, \citenamefont {Holleville}, \citenamefont {Gauguet},
  \citenamefont {Fils}, \citenamefont {Virdis}, \citenamefont {Clairon},
  \citenamefont {Dimarcq}, \citenamefont {Bord{\'e}}, \citenamefont {Landragin}
  \emph {et~al.}}]{I18}%
  \BibitemOpen
  \bibfield  {author} {\bibinfo {author} {\bibfnamefont {B.}~\bibnamefont
  {Canuel}}, \bibinfo {author} {\bibfnamefont {F.}~\bibnamefont {Leduc}},
  \bibinfo {author} {\bibfnamefont {D.}~\bibnamefont {Holleville}}, \bibinfo
  {author} {\bibfnamefont {A.}~\bibnamefont {Gauguet}}, \bibinfo {author}
  {\bibfnamefont {J.}~\bibnamefont {Fils}}, \bibinfo {author} {\bibfnamefont
  {A.}~\bibnamefont {Virdis}}, \bibinfo {author} {\bibfnamefont
  {A.}~\bibnamefont {Clairon}}, \bibinfo {author} {\bibfnamefont
  {N.}~\bibnamefont {Dimarcq}}, \bibinfo {author} {\bibfnamefont {C.~J.}\
  \bibnamefont {Bord{\'e}}}, \bibinfo {author} {\bibfnamefont {A.}~\bibnamefont
  {Landragin}}, \emph {et~al.},\ }\bibfield  {title} {\bibinfo {title}
  {Six-axis inertial sensor using cold-atom interferometry},\ }\href@noop {}
  {\bibfield  {journal} {\bibinfo  {journal} {Physical review letters}\
  }\textbf {\bibinfo {volume} {97}},\ \bibinfo {pages} {010402} (\bibinfo
  {year} {2006})}\BibitemShut {NoStop}%
\bibitem [{\citenamefont {McDonald}\ \emph {et~al.}(2013)\citenamefont
  {McDonald}, \citenamefont {Keal}, \citenamefont {Altin}, \citenamefont
  {Debs}, \citenamefont {Bennetts}, \citenamefont {Kuhn}, \citenamefont
  {Hardman}, \citenamefont {Johnsson}, \citenamefont {Close},\ and\
  \citenamefont {Robins}}]{I19}%
  \BibitemOpen
  \bibfield  {author} {\bibinfo {author} {\bibfnamefont {G.}~\bibnamefont
  {McDonald}}, \bibinfo {author} {\bibfnamefont {H.}~\bibnamefont {Keal}},
  \bibinfo {author} {\bibfnamefont {P.}~\bibnamefont {Altin}}, \bibinfo
  {author} {\bibfnamefont {J.}~\bibnamefont {Debs}}, \bibinfo {author}
  {\bibfnamefont {S.}~\bibnamefont {Bennetts}}, \bibinfo {author}
  {\bibfnamefont {C.}~\bibnamefont {Kuhn}}, \bibinfo {author} {\bibfnamefont
  {K.}~\bibnamefont {Hardman}}, \bibinfo {author} {\bibfnamefont
  {M.}~\bibnamefont {Johnsson}}, \bibinfo {author} {\bibfnamefont
  {J.}~\bibnamefont {Close}},\ and\ \bibinfo {author} {\bibfnamefont
  {N.}~\bibnamefont {Robins}},\ }\bibfield  {title} {\bibinfo {title}
  {Optically guided linear mach-zehnder atom interferometer},\ }\href@noop {}
  {\bibfield  {journal} {\bibinfo  {journal} {Physical Review A}\ }\textbf
  {\bibinfo {volume} {87}},\ \bibinfo {pages} {013632} (\bibinfo {year}
  {2013})}\BibitemShut {NoStop}%
\bibitem [{\citenamefont {Lenef}\ \emph {et~al.}(1997)\citenamefont {Lenef},
  \citenamefont {Hammond}, \citenamefont {Smith}, \citenamefont {Chapman},
  \citenamefont {Rubenstein},\ and\ \citenamefont {Pritchard}}]{I20}%
  \BibitemOpen
  \bibfield  {author} {\bibinfo {author} {\bibfnamefont {A.}~\bibnamefont
  {Lenef}}, \bibinfo {author} {\bibfnamefont {T.~D.}\ \bibnamefont {Hammond}},
  \bibinfo {author} {\bibfnamefont {E.~T.}\ \bibnamefont {Smith}}, \bibinfo
  {author} {\bibfnamefont {M.~S.}\ \bibnamefont {Chapman}}, \bibinfo {author}
  {\bibfnamefont {R.~A.}\ \bibnamefont {Rubenstein}},\ and\ \bibinfo {author}
  {\bibfnamefont {D.~E.}\ \bibnamefont {Pritchard}},\ }\bibfield  {title}
  {\bibinfo {title} {Rotation sensing with an atom interferometer},\
  }\href@noop {} {\bibfield  {journal} {\bibinfo  {journal} {Physical review
  letters}\ }\textbf {\bibinfo {volume} {78}},\ \bibinfo {pages} {760}
  (\bibinfo {year} {1997})}\BibitemShut {NoStop}%
\bibitem [{\citenamefont {Gustavson}\ \emph {et~al.}(1997)\citenamefont
  {Gustavson}, \citenamefont {Bouyer},\ and\ \citenamefont {Kasevich}}]{I21}%
  \BibitemOpen
  \bibfield  {author} {\bibinfo {author} {\bibfnamefont {T.}~\bibnamefont
  {Gustavson}}, \bibinfo {author} {\bibfnamefont {P.}~\bibnamefont {Bouyer}},\
  and\ \bibinfo {author} {\bibfnamefont {M.}~\bibnamefont {Kasevich}},\
  }\bibfield  {title} {\bibinfo {title} {Precision rotation measurements with
  an atom interferometer gyroscope},\ }\href@noop {} {\bibfield  {journal}
  {\bibinfo  {journal} {Physical review letters}\ }\textbf {\bibinfo {volume}
  {78}},\ \bibinfo {pages} {2046} (\bibinfo {year} {1997})}\BibitemShut
  {NoStop}%
\bibitem [{\citenamefont {Kasevich}\ and\ \citenamefont {Chu}(1992)}]{I22}%
  \BibitemOpen
  \bibfield  {author} {\bibinfo {author} {\bibfnamefont {M.}~\bibnamefont
  {Kasevich}}\ and\ \bibinfo {author} {\bibfnamefont {S.}~\bibnamefont {Chu}},\
  }\bibfield  {title} {\bibinfo {title} {Measurement of the gravitational
  acceleration of an atom with a light-pulse atom interferometer},\ }\href@noop
  {} {\bibfield  {journal} {\bibinfo  {journal} {Applied Physics B}\ }\textbf
  {\bibinfo {volume} {54}},\ \bibinfo {pages} {321} (\bibinfo {year}
  {1992})}\BibitemShut {NoStop}%
\bibitem [{\citenamefont {Peters}\ \emph {et~al.}(1999)\citenamefont {Peters},
  \citenamefont {Chung},\ and\ \citenamefont {Chu}}]{I23}%
  \BibitemOpen
  \bibfield  {author} {\bibinfo {author} {\bibfnamefont {A.}~\bibnamefont
  {Peters}}, \bibinfo {author} {\bibfnamefont {K.~Y.}\ \bibnamefont {Chung}},\
  and\ \bibinfo {author} {\bibfnamefont {S.}~\bibnamefont {Chu}},\ }\bibfield
  {title} {\bibinfo {title} {Measurement of gravitational acceleration by
  dropping atoms},\ }\href@noop {} {\bibfield  {journal} {\bibinfo  {journal}
  {Nature}\ }\textbf {\bibinfo {volume} {400}},\ \bibinfo {pages} {849}
  (\bibinfo {year} {1999})}\BibitemShut {NoStop}%
\bibitem [{\citenamefont {Geiger}\ \emph {et~al.}(2011)\citenamefont {Geiger},
  \citenamefont {M{\'e}noret}, \citenamefont {Stern}, \citenamefont {Zahzam},
  \citenamefont {Cheinet}, \citenamefont {Battelier}, \citenamefont {Villing},
  \citenamefont {Moron}, \citenamefont {Lours}, \citenamefont {Bidel} \emph
  {et~al.}}]{I24}%
  \BibitemOpen
  \bibfield  {author} {\bibinfo {author} {\bibfnamefont {R.}~\bibnamefont
  {Geiger}}, \bibinfo {author} {\bibfnamefont {V.}~\bibnamefont {M{\'e}noret}},
  \bibinfo {author} {\bibfnamefont {G.}~\bibnamefont {Stern}}, \bibinfo
  {author} {\bibfnamefont {N.}~\bibnamefont {Zahzam}}, \bibinfo {author}
  {\bibfnamefont {P.}~\bibnamefont {Cheinet}}, \bibinfo {author} {\bibfnamefont
  {B.}~\bibnamefont {Battelier}}, \bibinfo {author} {\bibfnamefont
  {A.}~\bibnamefont {Villing}}, \bibinfo {author} {\bibfnamefont
  {F.}~\bibnamefont {Moron}}, \bibinfo {author} {\bibfnamefont
  {M.}~\bibnamefont {Lours}}, \bibinfo {author} {\bibfnamefont
  {Y.}~\bibnamefont {Bidel}}, \emph {et~al.},\ }\bibfield  {title} {\bibinfo
  {title} {Detecting inertial effects with airborne matter-wave
  interferometry},\ }\href@noop {} {\bibfield  {journal} {\bibinfo  {journal}
  {Nature communications}\ }\textbf {\bibinfo {volume} {2}},\ \bibinfo {pages}
  {474} (\bibinfo {year} {2011})}\BibitemShut {NoStop}%
\bibitem [{\citenamefont {Snadden}\ \emph {et~al.}(1998)\citenamefont
  {Snadden}, \citenamefont {McGuirk}, \citenamefont {Bouyer}, \citenamefont
  {Haritos},\ and\ \citenamefont {Kasevich}}]{I25}%
  \BibitemOpen
  \bibfield  {author} {\bibinfo {author} {\bibfnamefont {M.}~\bibnamefont
  {Snadden}}, \bibinfo {author} {\bibfnamefont {J.}~\bibnamefont {McGuirk}},
  \bibinfo {author} {\bibfnamefont {P.}~\bibnamefont {Bouyer}}, \bibinfo
  {author} {\bibfnamefont {K.}~\bibnamefont {Haritos}},\ and\ \bibinfo {author}
  {\bibfnamefont {M.}~\bibnamefont {Kasevich}},\ }\bibfield  {title} {\bibinfo
  {title} {Measurement of the earth's gravity gradient with an atom
  interferometer-based gravity gradiometer},\ }\href@noop {} {\bibfield
  {journal} {\bibinfo  {journal} {Physical review letters}\ }\textbf {\bibinfo
  {volume} {81}},\ \bibinfo {pages} {971} (\bibinfo {year} {1998})}\BibitemShut
  {NoStop}%
\bibitem [{\citenamefont {Mcguirk}\ \emph {et~al.}(2002)\citenamefont
  {Mcguirk}, \citenamefont {Foster}, \citenamefont {Fixler}, \citenamefont
  {Snadden},\ and\ \citenamefont {Kasevich}}]{I26}%
  \BibitemOpen
  \bibfield  {author} {\bibinfo {author} {\bibfnamefont {J.~M.}\ \bibnamefont
  {Mcguirk}}, \bibinfo {author} {\bibfnamefont {G.}~\bibnamefont {Foster}},
  \bibinfo {author} {\bibfnamefont {J.}~\bibnamefont {Fixler}}, \bibinfo
  {author} {\bibfnamefont {M.}~\bibnamefont {Snadden}},\ and\ \bibinfo {author}
  {\bibfnamefont {M.}~\bibnamefont {Kasevich}},\ }\bibfield  {title} {\bibinfo
  {title} {Sensitive absolute-gravity gradiometry using atom interferometry},\
  }\href@noop {} {\bibfield  {journal} {\bibinfo  {journal} {Physical Review
  A}\ }\textbf {\bibinfo {volume} {65}},\ \bibinfo {pages} {033608} (\bibinfo
  {year} {2002})}\BibitemShut {NoStop}%
\bibitem [{\citenamefont {Lauterbur}\ \emph {et~al.}(1973)\citenamefont
  {Lauterbur} \emph {et~al.}}]{I27}%
  \BibitemOpen
  \bibfield  {author} {\bibinfo {author} {\bibfnamefont {P.~C.}\ \bibnamefont
  {Lauterbur}} \emph {et~al.},\ }\bibfield  {title} {\bibinfo {title} {Image
  formation by induced local interactions: examples employing nuclear magnetic
  resonance},\ }\href@noop {} {\  (\bibinfo {year} {1973})}\BibitemShut
  {NoStop}%
\bibitem [{\citenamefont {Griffin}\ \emph {et~al.}(1996)\citenamefont
  {Griffin}, \citenamefont {Snoke},\ and\ \citenamefont {Stringari}}]{3_1}%
  \BibitemOpen
  \bibfield  {author} {\bibinfo {author} {\bibfnamefont {A.}~\bibnamefont
  {Griffin}}, \bibinfo {author} {\bibfnamefont {D.~W.}\ \bibnamefont {Snoke}},\
  and\ \bibinfo {author} {\bibfnamefont {S.}~\bibnamefont {Stringari}},\
  }\href@noop {} {\emph {\bibinfo {title} {Bose-einstein condensation}}}\
  (\bibinfo  {publisher} {Cambridge University Press},\ \bibinfo {year}
  {1996})\BibitemShut {NoStop}%
\bibitem [{\citenamefont {Timp}\ \emph {et~al.}(1992)\citenamefont {Timp},
  \citenamefont {Behringer}, \citenamefont {Tennant}, \citenamefont
  {Cunningham}, \citenamefont {Prentiss},\ and\ \citenamefont {Berggren}}]{p1}%
  \BibitemOpen
  \bibfield  {author} {\bibinfo {author} {\bibfnamefont {G.}~\bibnamefont
  {Timp}}, \bibinfo {author} {\bibfnamefont {R.}~\bibnamefont {Behringer}},
  \bibinfo {author} {\bibfnamefont {D.}~\bibnamefont {Tennant}}, \bibinfo
  {author} {\bibfnamefont {J.}~\bibnamefont {Cunningham}}, \bibinfo {author}
  {\bibfnamefont {M.}~\bibnamefont {Prentiss}},\ and\ \bibinfo {author}
  {\bibfnamefont {K.}~\bibnamefont {Berggren}},\ }\bibfield  {title} {\bibinfo
  {title} {Using light as a lens for submicron, neutral-atom lithography},\
  }\href@noop {} {\bibfield  {journal} {\bibinfo  {journal} {Physical review
  letters}\ }\textbf {\bibinfo {volume} {69}},\ \bibinfo {pages} {1636}
  (\bibinfo {year} {1992})}\BibitemShut {NoStop}%
\bibitem [{\citenamefont {McClelland}\ \emph {et~al.}(1993)\citenamefont
  {McClelland}, \citenamefont {Scholten}, \citenamefont {Palm},\ and\
  \citenamefont {Celotta}}]{p2}%
  \BibitemOpen
  \bibfield  {author} {\bibinfo {author} {\bibfnamefont {J.~J.}\ \bibnamefont
  {McClelland}}, \bibinfo {author} {\bibfnamefont {R.}~\bibnamefont
  {Scholten}}, \bibinfo {author} {\bibfnamefont {E.}~\bibnamefont {Palm}},\
  and\ \bibinfo {author} {\bibfnamefont {R.~J.}\ \bibnamefont {Celotta}},\
  }\bibfield  {title} {\bibinfo {title} {Laser-focused atomic deposition},\
  }\href@noop {} {\bibfield  {journal} {\bibinfo  {journal} {Science}\ }\textbf
  {\bibinfo {volume} {262}},\ \bibinfo {pages} {877} (\bibinfo {year}
  {1993})}\BibitemShut {NoStop}%
\bibitem [{\citenamefont {Oberthaler}\ and\ \citenamefont {Pfau}(2003)}]{p49}%
  \BibitemOpen
  \bibfield  {author} {\bibinfo {author} {\bibfnamefont {M.~K.}\ \bibnamefont
  {Oberthaler}}\ and\ \bibinfo {author} {\bibfnamefont {T.}~\bibnamefont
  {Pfau}},\ }\bibfield  {title} {\bibinfo {title} {One-, two-and
  three-dimensional nanostructures with atom lithography},\ }\href@noop {}
  {\bibfield  {journal} {\bibinfo  {journal} {Journal of Physics: Condensed
  Matter}\ }\textbf {\bibinfo {volume} {15}},\ \bibinfo {pages} {R233}
  (\bibinfo {year} {2003})}\BibitemShut {NoStop}%
\bibitem [{\citenamefont {Ohmukai}\ \emph {et~al.}(2003)\citenamefont
  {Ohmukai}, \citenamefont {Urabe},\ and\ \citenamefont {Watanabe}}]{p50}%
  \BibitemOpen
  \bibfield  {author} {\bibinfo {author} {\bibfnamefont {R.}~\bibnamefont
  {Ohmukai}}, \bibinfo {author} {\bibfnamefont {S.}~\bibnamefont {Urabe}},\
  and\ \bibinfo {author} {\bibfnamefont {M.}~\bibnamefont {Watanabe}},\
  }\bibfield  {title} {\bibinfo {title} {Atom lithography with ytterbium
  beam},\ }\href@noop {} {\bibfield  {journal} {\bibinfo  {journal} {Applied
  Physics B}\ }\textbf {\bibinfo {volume} {77}},\ \bibinfo {pages} {415}
  (\bibinfo {year} {2003})}\BibitemShut {NoStop}%
\bibitem [{\citenamefont {Myszkiewicz}\ \emph {et~al.}(2004)\citenamefont
  {Myszkiewicz}, \citenamefont {Hohlfeld}, \citenamefont {Toonen},
  \citenamefont {Van~Etteger}, \citenamefont {Shklyarevskii}, \citenamefont
  {Meerts}, \citenamefont {Rasing},\ and\ \citenamefont {Jurdik}}]{p51}%
  \BibitemOpen
  \bibfield  {author} {\bibinfo {author} {\bibfnamefont {G.}~\bibnamefont
  {Myszkiewicz}}, \bibinfo {author} {\bibfnamefont {J.}~\bibnamefont
  {Hohlfeld}}, \bibinfo {author} {\bibfnamefont {A.}~\bibnamefont {Toonen}},
  \bibinfo {author} {\bibfnamefont {A.}~\bibnamefont {Van~Etteger}}, \bibinfo
  {author} {\bibfnamefont {O.}~\bibnamefont {Shklyarevskii}}, \bibinfo {author}
  {\bibfnamefont {W.}~\bibnamefont {Meerts}}, \bibinfo {author} {\bibfnamefont
  {T.}~\bibnamefont {Rasing}},\ and\ \bibinfo {author} {\bibfnamefont
  {E.}~\bibnamefont {Jurdik}},\ }\bibfield  {title} {\bibinfo {title} {Laser
  manipulation of iron for nanofabrication},\ }\href@noop {} {\bibfield
  {journal} {\bibinfo  {journal} {Applied physics letters}\ }\textbf {\bibinfo
  {volume} {85}},\ \bibinfo {pages} {3842} (\bibinfo {year}
  {2004})}\BibitemShut {NoStop}%
\bibitem [{\citenamefont {Smeets}\ \emph {et~al.}(2010)\citenamefont {Smeets},
  \citenamefont {van~der Straten}, \citenamefont {Meijer}, \citenamefont
  {Fabrie},\ and\ \citenamefont {van Leeuwen}}]{p52}%
  \BibitemOpen
  \bibfield  {author} {\bibinfo {author} {\bibfnamefont {B.}~\bibnamefont
  {Smeets}}, \bibinfo {author} {\bibfnamefont {P.}~\bibnamefont {van~der
  Straten}}, \bibinfo {author} {\bibfnamefont {T.}~\bibnamefont {Meijer}},
  \bibinfo {author} {\bibfnamefont {C.}~\bibnamefont {Fabrie}},\ and\ \bibinfo
  {author} {\bibfnamefont {K.}~\bibnamefont {van Leeuwen}},\ }\bibfield
  {title} {\bibinfo {title} {Atom lithography without laser cooling},\
  }\href@noop {} {\bibfield  {journal} {\bibinfo  {journal} {Applied Physics
  B}\ }\textbf {\bibinfo {volume} {98}},\ \bibinfo {pages} {697} (\bibinfo
  {year} {2010})}\BibitemShut {NoStop}%
\bibitem [{\citenamefont {Wiseman}(1997)}]{5_1}%
  \BibitemOpen
  \bibfield  {author} {\bibinfo {author} {\bibfnamefont {H.}~\bibnamefont
  {Wiseman}},\ }\bibfield  {title} {\bibinfo {title} {Defining the (atom)
  laser},\ }\href@noop {} {\bibfield  {journal} {\bibinfo  {journal} {Physical
  Review A}\ }\textbf {\bibinfo {volume} {56}},\ \bibinfo {pages} {2068}
  (\bibinfo {year} {1997})}\BibitemShut {NoStop}%
\bibitem [{\citenamefont {Bolpasi}\ \emph {et~al.}(2014)\citenamefont
  {Bolpasi}, \citenamefont {Efremidis}, \citenamefont {Morrissey},
  \citenamefont {Condylis}, \citenamefont {Sahagun}, \citenamefont {Baker},\
  and\ \citenamefont {Von~Klitzing}}]{5_32}%
  \BibitemOpen
  \bibfield  {author} {\bibinfo {author} {\bibfnamefont {V.}~\bibnamefont
  {Bolpasi}}, \bibinfo {author} {\bibfnamefont {N.}~\bibnamefont {Efremidis}},
  \bibinfo {author} {\bibfnamefont {M.}~\bibnamefont {Morrissey}}, \bibinfo
  {author} {\bibfnamefont {P.}~\bibnamefont {Condylis}}, \bibinfo {author}
  {\bibfnamefont {D.}~\bibnamefont {Sahagun}}, \bibinfo {author} {\bibfnamefont
  {M.}~\bibnamefont {Baker}},\ and\ \bibinfo {author} {\bibfnamefont
  {W.}~\bibnamefont {Von~Klitzing}},\ }\bibfield  {title} {\bibinfo {title} {An
  ultra-bright atom laser},\ }\href@noop {} {\bibfield  {journal} {\bibinfo
  {journal} {New Journal of Physics}\ }\textbf {\bibinfo {volume} {16}},\
  \bibinfo {pages} {033036} (\bibinfo {year} {2014})}\BibitemShut {NoStop}%
\bibitem [{\citenamefont {Richberg}\ \emph {et~al.}(2021)\citenamefont
  {Richberg}, \citenamefont {Szigeti},\ and\ \citenamefont {Martin}}]{Ri_3}%
  \BibitemOpen
  \bibfield  {author} {\bibinfo {author} {\bibfnamefont {R.}~\bibnamefont
  {Richberg}}, \bibinfo {author} {\bibfnamefont {S.}~\bibnamefont {Szigeti}},\
  and\ \bibinfo {author} {\bibfnamefont {A.}~\bibnamefont {Martin}},\
  }\bibfield  {title} {\bibinfo {title} {Optical focusing of bose-einstein
  condensates},\ }\href@noop {} {\bibfield  {journal} {\bibinfo  {journal}
  {Physical Review A}\ }\textbf {\bibinfo {volume} {103}},\ \bibinfo {pages}
  {063304} (\bibinfo {year} {2021})}\BibitemShut {NoStop}%
\bibitem [{\citenamefont {Richberg}\ and\ \citenamefont {Martin}(2021)}]{Ri_4}%
  \BibitemOpen
  \bibfield  {author} {\bibinfo {author} {\bibfnamefont {R.}~\bibnamefont
  {Richberg}}\ and\ \bibinfo {author} {\bibfnamefont {A.~M.}\ \bibnamefont
  {Martin}},\ }\bibfield  {title} {\bibinfo {title} {The influence of s-wave
  interactions on focussing of atoms},\ }\bibfield  {journal} {\bibinfo
  {journal} {Atoms}\ }\textbf {\bibinfo {volume} {9}},\ \href
  {https://doi.org/10.3390/atoms9030037} {10.3390/atoms9030037} (\bibinfo
  {year} {2021})\BibitemShut {NoStop}%
\bibitem [{\citenamefont {Kordbacheh}\ and\ \citenamefont
  {Martin}(2020)}]{Ri_1}%
  \BibitemOpen
  \bibfield  {author} {\bibinfo {author} {\bibfnamefont {A.}~\bibnamefont
  {Kordbacheh}}\ and\ \bibinfo {author} {\bibfnamefont {A.}~\bibnamefont
  {Martin}},\ }\bibfield  {title} {\bibinfo {title} {The influence of s-wave
  interactions on focussing of atoms},\ }\href@noop {} {\bibfield  {journal}
  {\bibinfo  {journal} {arXiv preprint arXiv:2012.04892}\ } (\bibinfo {year}
  {2020})}\BibitemShut {NoStop}%
\bibitem [{\citenamefont {Kordbacheh}\ \emph {et~al.}(2020)\citenamefont
  {Kordbacheh}, \citenamefont {Szigeti},\ and\ \citenamefont {Martin}}]{Ri_2}%
  \BibitemOpen
  \bibfield  {author} {\bibinfo {author} {\bibfnamefont {A.}~\bibnamefont
  {Kordbacheh}}, \bibinfo {author} {\bibfnamefont {S.}~\bibnamefont
  {Szigeti}},\ and\ \bibinfo {author} {\bibfnamefont {A.}~\bibnamefont
  {Martin}},\ }\bibfield  {title} {\bibinfo {title} {Optical focusing of
  bose-einstein condensates},\ }\href@noop {} {\bibfield  {journal} {\bibinfo
  {journal} {arXiv preprint arXiv:2011.14470}\ } (\bibinfo {year}
  {2020})}\BibitemShut {NoStop}%
\bibitem [{\citenamefont {Riou}\ \emph {et~al.}(2008)\citenamefont {Riou},
  \citenamefont {Le~Coq}, \citenamefont {Impens}, \citenamefont {Guerin},
  \citenamefont {Bord{\'e}}, \citenamefont {Aspect},\ and\ \citenamefont
  {Bouyer}}]{AL_1}%
  \BibitemOpen
  \bibfield  {author} {\bibinfo {author} {\bibfnamefont {J.-F.}\ \bibnamefont
  {Riou}}, \bibinfo {author} {\bibfnamefont {Y.}~\bibnamefont {Le~Coq}},
  \bibinfo {author} {\bibfnamefont {F.}~\bibnamefont {Impens}}, \bibinfo
  {author} {\bibfnamefont {W.}~\bibnamefont {Guerin}}, \bibinfo {author}
  {\bibfnamefont {C.}~\bibnamefont {Bord{\'e}}}, \bibinfo {author}
  {\bibfnamefont {A.}~\bibnamefont {Aspect}},\ and\ \bibinfo {author}
  {\bibfnamefont {P.}~\bibnamefont {Bouyer}},\ }\bibfield  {title} {\bibinfo
  {title} {Theoretical tools for atom-laser-beam propagation},\ }\href@noop {}
  {\bibfield  {journal} {\bibinfo  {journal} {Physical Review A}\ }\textbf
  {\bibinfo {volume} {77}},\ \bibinfo {pages} {033630} (\bibinfo {year}
  {2008})}\BibitemShut {NoStop}%
\bibitem [{\citenamefont {Impens}\ and\ \citenamefont
  {Bord{\'e}}(2009)}]{AL_2}%
  \BibitemOpen
  \bibfield  {author} {\bibinfo {author} {\bibfnamefont {F.}~\bibnamefont
  {Impens}}\ and\ \bibinfo {author} {\bibfnamefont {C.~J.}\ \bibnamefont
  {Bord{\'e}}},\ }\bibfield  {title} {\bibinfo {title} {Generalized a b c d
  propagation for interacting atomic clouds},\ }\href@noop {} {\bibfield
  {journal} {\bibinfo  {journal} {Physical Review A}\ }\textbf {\bibinfo
  {volume} {79}},\ \bibinfo {pages} {043613} (\bibinfo {year}
  {2009})}\BibitemShut {NoStop}%
\bibitem [{\citenamefont {Impens}(2009)}]{AL_3}%
  \BibitemOpen
  \bibfield  {author} {\bibinfo {author} {\bibfnamefont {F.}~\bibnamefont
  {Impens}},\ }\bibfield  {title} {\bibinfo {title} {Hidden symmetry and
  nonlinear paraxial atom optics},\ }\href@noop {} {\bibfield  {journal}
  {\bibinfo  {journal} {Physical Review A}\ }\textbf {\bibinfo {volume} {80}},\
  \bibinfo {pages} {063617} (\bibinfo {year} {2009})}\BibitemShut {NoStop}%
\bibitem [{\citenamefont {Mewes}\ \emph {et~al.}(1997)\citenamefont {Mewes},
  \citenamefont {Andrews}, \citenamefont {Kurn}, \citenamefont {Durfee},
  \citenamefont {Townsend},\ and\ \citenamefont {Ketterle}}]{5_6}%
  \BibitemOpen
  \bibfield  {author} {\bibinfo {author} {\bibfnamefont {M.-O.}\ \bibnamefont
  {Mewes}}, \bibinfo {author} {\bibfnamefont {M.}~\bibnamefont {Andrews}},
  \bibinfo {author} {\bibfnamefont {D.}~\bibnamefont {Kurn}}, \bibinfo {author}
  {\bibfnamefont {D.}~\bibnamefont {Durfee}}, \bibinfo {author} {\bibfnamefont
  {C.}~\bibnamefont {Townsend}},\ and\ \bibinfo {author} {\bibfnamefont
  {W.}~\bibnamefont {Ketterle}},\ }\bibfield  {title} {\bibinfo {title} {Output
  coupler for bose-einstein condensed atoms},\ }\href@noop {} {\bibfield
  {journal} {\bibinfo  {journal} {Physical Review Letters}\ }\textbf {\bibinfo
  {volume} {78}},\ \bibinfo {pages} {582} (\bibinfo {year} {1997})}\BibitemShut
  {NoStop}%
\bibitem [{\citenamefont {Riou}\ \emph {et~al.}(2006)\citenamefont {Riou},
  \citenamefont {Guerin}, \citenamefont {Le~Coq}, \citenamefont
  {Fauquembergue}, \citenamefont {Josse}, \citenamefont {Bouyer},\ and\
  \citenamefont {Aspect}}]{5_13}%
  \BibitemOpen
  \bibfield  {author} {\bibinfo {author} {\bibfnamefont {J.-F.}\ \bibnamefont
  {Riou}}, \bibinfo {author} {\bibfnamefont {W.}~\bibnamefont {Guerin}},
  \bibinfo {author} {\bibfnamefont {Y.}~\bibnamefont {Le~Coq}}, \bibinfo
  {author} {\bibfnamefont {M.}~\bibnamefont {Fauquembergue}}, \bibinfo {author}
  {\bibfnamefont {V.}~\bibnamefont {Josse}}, \bibinfo {author} {\bibfnamefont
  {P.}~\bibnamefont {Bouyer}},\ and\ \bibinfo {author} {\bibfnamefont
  {A.}~\bibnamefont {Aspect}},\ }\bibfield  {title} {\bibinfo {title} {Beam
  quality of a nonideal atom laser},\ }\href@noop {} {\bibfield  {journal}
  {\bibinfo  {journal} {Physical review letters}\ }\textbf {\bibinfo {volume}
  {96}},\ \bibinfo {pages} {070404} (\bibinfo {year} {2006})}\BibitemShut
  {NoStop}%
\bibitem [{\citenamefont {Hagley}\ \emph {et~al.}(1999)\citenamefont {Hagley},
  \citenamefont {Deng}, \citenamefont {Kozuma}, \citenamefont {Wen},
  \citenamefont {Helmerson}, \citenamefont {Rolston}, ,\ and\ \citenamefont
  {Phillips}}]{5_7}%
  \BibitemOpen
  \bibfield  {author} {\bibinfo {author} {\bibfnamefont {E.~W.}\ \bibnamefont
  {Hagley}}, \bibinfo {author} {\bibfnamefont {L.}~\bibnamefont {Deng}},
  \bibinfo {author} {\bibfnamefont {M.}~\bibnamefont {Kozuma}}, \bibinfo
  {author} {\bibfnamefont {J.}~\bibnamefont {Wen}}, \bibinfo {author}
  {\bibfnamefont {K.}~\bibnamefont {Helmerson}}, \bibinfo {author}
  {\bibfnamefont {S.}~\bibnamefont {Rolston}}, ,\ and\ \bibinfo {author}
  {\bibfnamefont {W.~D.}\ \bibnamefont {Phillips}},\ }\bibfield  {title}
  {\bibinfo {title} {A well-collimated quasi-continuous atom laser},\
  }\href@noop {} {\bibfield  {journal} {\bibinfo  {journal} {Science}\ }\textbf
  {\bibinfo {volume} {283}},\ \bibinfo {pages} {1706} (\bibinfo {year}
  {1999})}\BibitemShut {NoStop}%
\bibitem [{\citenamefont {Bloch}\ \emph {et~al.}(1999)\citenamefont {Bloch},
  \citenamefont {H{\"a}nsch},\ and\ \citenamefont {Esslinger}}]{5_9}%
  \BibitemOpen
  \bibfield  {author} {\bibinfo {author} {\bibfnamefont {I.}~\bibnamefont
  {Bloch}}, \bibinfo {author} {\bibfnamefont {T.~W.}\ \bibnamefont
  {H{\"a}nsch}},\ and\ \bibinfo {author} {\bibfnamefont {T.}~\bibnamefont
  {Esslinger}},\ }\bibfield  {title} {\bibinfo {title} {Atom laser with a cw
  output coupler},\ }\href@noop {} {\bibfield  {journal} {\bibinfo  {journal}
  {Physical Review Letters}\ }\textbf {\bibinfo {volume} {82}},\ \bibinfo
  {pages} {3008} (\bibinfo {year} {1999})}\BibitemShut {NoStop}%
\bibitem [{\citenamefont {Esslinger}\ \emph {et~al.}(1998)\citenamefont
  {Esslinger}, \citenamefont {Bloch},\ and\ \citenamefont {H{\"a}nsch}}]{5_10}%
  \BibitemOpen
  \bibfield  {author} {\bibinfo {author} {\bibfnamefont {T.}~\bibnamefont
  {Esslinger}}, \bibinfo {author} {\bibfnamefont {I.}~\bibnamefont {Bloch}},\
  and\ \bibinfo {author} {\bibfnamefont {T.~W.}\ \bibnamefont {H{\"a}nsch}},\
  }\bibfield  {title} {\bibinfo {title} {Bose-einstein condensation in a
  quadrupole-ioffe-configuration trap},\ }\href@noop {} {\bibfield  {journal}
  {\bibinfo  {journal} {Physical Review A}\ }\textbf {\bibinfo {volume} {58}},\
  \bibinfo {pages} {R2664} (\bibinfo {year} {1998})}\BibitemShut {NoStop}%
\bibitem [{\citenamefont {Robins}\ \emph {et~al.}(2006)\citenamefont {Robins},
  \citenamefont {Figl}, \citenamefont {Haine}, \citenamefont {Morrison},
  \citenamefont {Jeppesen}, \citenamefont {Hope},\ and\ \citenamefont
  {Close}}]{5_11}%
  \BibitemOpen
  \bibfield  {author} {\bibinfo {author} {\bibfnamefont {N.}~\bibnamefont
  {Robins}}, \bibinfo {author} {\bibfnamefont {C.}~\bibnamefont {Figl}},
  \bibinfo {author} {\bibfnamefont {S.}~\bibnamefont {Haine}}, \bibinfo
  {author} {\bibfnamefont {A.}~\bibnamefont {Morrison}}, \bibinfo {author}
  {\bibfnamefont {M.}~\bibnamefont {Jeppesen}}, \bibinfo {author}
  {\bibfnamefont {J.}~\bibnamefont {Hope}},\ and\ \bibinfo {author}
  {\bibfnamefont {J.}~\bibnamefont {Close}},\ }\bibfield  {title} {\bibinfo
  {title} {Achieving peak brightness in an atom laser},\ }\href@noop {}
  {\bibfield  {journal} {\bibinfo  {journal} {Physical review letters}\
  }\textbf {\bibinfo {volume} {96}},\ \bibinfo {pages} {140403} (\bibinfo
  {year} {2006})}\BibitemShut {NoStop}%
\bibitem [{\citenamefont {Moy}\ \emph {et~al.}(1997)\citenamefont {Moy},
  \citenamefont {Hope},\ and\ \citenamefont {Savage}}]{5_12}%
  \BibitemOpen
  \bibfield  {author} {\bibinfo {author} {\bibfnamefont {G.}~\bibnamefont
  {Moy}}, \bibinfo {author} {\bibfnamefont {J.}~\bibnamefont {Hope}},\ and\
  \bibinfo {author} {\bibfnamefont {C.}~\bibnamefont {Savage}},\ }\bibfield
  {title} {\bibinfo {title} {Atom laser based on raman transitions},\
  }\href@noop {} {\bibfield  {journal} {\bibinfo  {journal} {Physical Review
  A}\ }\textbf {\bibinfo {volume} {55}},\ \bibinfo {pages} {3631} (\bibinfo
  {year} {1997})}\BibitemShut {NoStop}%
\bibitem [{\citenamefont {Le~Coq}\ \emph {et~al.}(2001)\citenamefont {Le~Coq},
  \citenamefont {Thywissen}, \citenamefont {Rangwala}, \citenamefont {Gerbier},
  \citenamefont {Richard}, \citenamefont {Delannoy}, \citenamefont {Bouyer},\
  and\ \citenamefont {Aspect}}]{AL_4}%
  \BibitemOpen
  \bibfield  {author} {\bibinfo {author} {\bibfnamefont {Y.}~\bibnamefont
  {Le~Coq}}, \bibinfo {author} {\bibfnamefont {J.~H.}\ \bibnamefont
  {Thywissen}}, \bibinfo {author} {\bibfnamefont {S.~A.}\ \bibnamefont
  {Rangwala}}, \bibinfo {author} {\bibfnamefont {F.}~\bibnamefont {Gerbier}},
  \bibinfo {author} {\bibfnamefont {S.}~\bibnamefont {Richard}}, \bibinfo
  {author} {\bibfnamefont {G.}~\bibnamefont {Delannoy}}, \bibinfo {author}
  {\bibfnamefont {P.}~\bibnamefont {Bouyer}},\ and\ \bibinfo {author}
  {\bibfnamefont {A.}~\bibnamefont {Aspect}},\ }\bibfield  {title} {\bibinfo
  {title} {Atom laser divergence},\ }\href@noop {} {\bibfield  {journal}
  {\bibinfo  {journal} {Physical review letters}\ }\textbf {\bibinfo {volume}
  {87}},\ \bibinfo {pages} {170403} (\bibinfo {year} {2001})}\BibitemShut
  {NoStop}%
\bibitem [{\citenamefont {Jeppesen}\ \emph {et~al.}(2008)\citenamefont
  {Jeppesen}, \citenamefont {Dugu{\'e}}, \citenamefont {Dennis}, \citenamefont
  {Johnsson}, \citenamefont {Figl}, \citenamefont {Robins},\ and\ \citenamefont
  {Close}}]{5_18}%
  \BibitemOpen
  \bibfield  {author} {\bibinfo {author} {\bibfnamefont {M.}~\bibnamefont
  {Jeppesen}}, \bibinfo {author} {\bibfnamefont {J.}~\bibnamefont {Dugu{\'e}}},
  \bibinfo {author} {\bibfnamefont {G.}~\bibnamefont {Dennis}}, \bibinfo
  {author} {\bibfnamefont {M.}~\bibnamefont {Johnsson}}, \bibinfo {author}
  {\bibfnamefont {C.}~\bibnamefont {Figl}}, \bibinfo {author} {\bibfnamefont
  {N.}~\bibnamefont {Robins}},\ and\ \bibinfo {author} {\bibfnamefont
  {J.}~\bibnamefont {Close}},\ }\bibfield  {title} {\bibinfo {title}
  {Approaching the heisenberg limit in an atom laser},\ }\href@noop {}
  {\bibfield  {journal} {\bibinfo  {journal} {Physical Review A}\ }\textbf
  {\bibinfo {volume} {77}},\ \bibinfo {pages} {063618} (\bibinfo {year}
  {2008})}\BibitemShut {NoStop}%
\bibitem [{\citenamefont {Edwards}\ \emph {et~al.}(1999)\citenamefont
  {Edwards}, \citenamefont {Griggs}, \citenamefont {Holman}, \citenamefont
  {Clark}, \citenamefont {Rolston},\ and\ \citenamefont {Phillips}}]{5_31}%
  \BibitemOpen
  \bibfield  {author} {\bibinfo {author} {\bibfnamefont {M.}~\bibnamefont
  {Edwards}}, \bibinfo {author} {\bibfnamefont {D.~A.}\ \bibnamefont {Griggs}},
  \bibinfo {author} {\bibfnamefont {P.~L.}\ \bibnamefont {Holman}}, \bibinfo
  {author} {\bibfnamefont {C.~W.}\ \bibnamefont {Clark}}, \bibinfo {author}
  {\bibfnamefont {S.}~\bibnamefont {Rolston}},\ and\ \bibinfo {author}
  {\bibfnamefont {W.~D.}\ \bibnamefont {Phillips}},\ }\bibfield  {title}
  {\bibinfo {title} {Properties of a raman atom-laser output coupler},\
  }\href@noop {} {\bibfield  {journal} {\bibinfo  {journal} {Journal of Physics
  B: Atomic, Molecular and Optical Physics}\ }\textbf {\bibinfo {volume}
  {32}},\ \bibinfo {pages} {2935} (\bibinfo {year} {1999})}\BibitemShut
  {NoStop}%
\bibitem [{\citenamefont {Gross}(1961)}]{3_8}%
  \BibitemOpen
  \bibfield  {author} {\bibinfo {author} {\bibfnamefont {E.~P.}\ \bibnamefont
  {Gross}},\ }\bibfield  {title} {\bibinfo {title} {Structure of a quantized
  vortex in boson systems},\ }\href@noop {} {\bibfield  {journal} {\bibinfo
  {journal} {Il Nuovo Cimento (1955-1965)}\ }\textbf {\bibinfo {volume} {20}},\
  \bibinfo {pages} {454} (\bibinfo {year} {1961})}\BibitemShut {NoStop}%
\bibitem [{\citenamefont {Pitaevskii}(1961)}]{3_9}%
  \BibitemOpen
  \bibfield  {author} {\bibinfo {author} {\bibfnamefont {L.}~\bibnamefont
  {Pitaevskii}},\ }\bibfield  {title} {\bibinfo {title} {Vortex lines in an
  imperfect bose gas},\ }\href@noop {} {\bibfield  {journal} {\bibinfo
  {journal} {Sov. Phys. JETP}\ }\textbf {\bibinfo {volume} {13}},\ \bibinfo
  {pages} {451} (\bibinfo {year} {1961})}\BibitemShut {NoStop}%
\bibitem [{\citenamefont {Stenger}\ \emph {et~al.}(1999)\citenamefont
  {Stenger}, \citenamefont {Inouye}, \citenamefont {Chikkatur}, \citenamefont
  {Stamper-Kurn}, \citenamefont {Pritchard},\ and\ \citenamefont
  {Ketterle}}]{5_8}%
  \BibitemOpen
  \bibfield  {author} {\bibinfo {author} {\bibfnamefont {J.}~\bibnamefont
  {Stenger}}, \bibinfo {author} {\bibfnamefont {S.}~\bibnamefont {Inouye}},
  \bibinfo {author} {\bibfnamefont {A.~P.}\ \bibnamefont {Chikkatur}}, \bibinfo
  {author} {\bibfnamefont {D.}~\bibnamefont {Stamper-Kurn}}, \bibinfo {author}
  {\bibfnamefont {D.}~\bibnamefont {Pritchard}},\ and\ \bibinfo {author}
  {\bibfnamefont {W.}~\bibnamefont {Ketterle}},\ }\bibfield  {title} {\bibinfo
  {title} {Bragg spectroscopy of a bose-einstein condensate},\ }\href@noop {}
  {\bibfield  {journal} {\bibinfo  {journal} {Physical Review Letters}\
  }\textbf {\bibinfo {volume} {82}},\ \bibinfo {pages} {4569} (\bibinfo {year}
  {1999})}\BibitemShut {NoStop}%
\bibitem [{\citenamefont {Dalfovo}\ \emph {et~al.}(1999)\citenamefont
  {Dalfovo}, \citenamefont {Giorgini}, \citenamefont {Pitaevskii},\ and\
  \citenamefont {Stringari}}]{10}%
  \BibitemOpen
  \bibfield  {author} {\bibinfo {author} {\bibfnamefont {F.}~\bibnamefont
  {Dalfovo}}, \bibinfo {author} {\bibfnamefont {S.}~\bibnamefont {Giorgini}},
  \bibinfo {author} {\bibfnamefont {L.~P.}\ \bibnamefont {Pitaevskii}},\ and\
  \bibinfo {author} {\bibfnamefont {S.}~\bibnamefont {Stringari}},\ }\bibfield
  {title} {\bibinfo {title} {Theory of bose-einstein condensation in trapped
  gases},\ }\href@noop {} {\bibfield  {journal} {\bibinfo  {journal} {Reviews
  of Modern Physics}\ }\textbf {\bibinfo {volume} {71}},\ \bibinfo {pages}
  {463} (\bibinfo {year} {1999})}\BibitemShut {NoStop}%
\bibitem [{\citenamefont {Pitaevskii}\ and\ \citenamefont
  {Stringari}(2003)}]{11}%
  \BibitemOpen
  \bibfield  {author} {\bibinfo {author} {\bibfnamefont {L.}~\bibnamefont
  {Pitaevskii}}\ and\ \bibinfo {author} {\bibfnamefont {S.}~\bibnamefont
  {Stringari}},\ }\href@noop {} {\bibinfo {title} {Bose-einstein condensation,
  clarendon}} (\bibinfo {year} {2003})\BibitemShut {NoStop}%
\bibitem [{\citenamefont {Pethick}\ and\ \citenamefont {Smith}(2002)}]{12}%
  \BibitemOpen
  \bibfield  {author} {\bibinfo {author} {\bibfnamefont {C.~J.}\ \bibnamefont
  {Pethick}}\ and\ \bibinfo {author} {\bibfnamefont {H.}~\bibnamefont
  {Smith}},\ }\href@noop {} {\emph {\bibinfo {title} {Bose-Einstein
  condensation in dilute gases}}}\ (\bibinfo  {publisher} {Cambridge university
  press},\ \bibinfo {year} {2002})\BibitemShut {NoStop}%
\bibitem [{\citenamefont {Bogoliubov}(1947)}]{13}%
  \BibitemOpen
  \bibfield  {author} {\bibinfo {author} {\bibfnamefont {N.}~\bibnamefont
  {Bogoliubov}},\ }\bibfield  {title} {\bibinfo {title} {On the theory of
  superfluidity},\ }\href@noop {} {\bibfield  {journal} {\bibinfo  {journal}
  {J. Phys}\ }\textbf {\bibinfo {volume} {11}},\ \bibinfo {pages} {23}
  (\bibinfo {year} {1947})}\BibitemShut {NoStop}%
\bibitem [{\citenamefont {Braaten}\ \emph {et~al.}(2002)\citenamefont
  {Braaten}, \citenamefont {Hammer},\ and\ \citenamefont {Mehen}}]{15}%
  \BibitemOpen
  \bibfield  {author} {\bibinfo {author} {\bibfnamefont {E.}~\bibnamefont
  {Braaten}}, \bibinfo {author} {\bibfnamefont {H.-W.}\ \bibnamefont
  {Hammer}},\ and\ \bibinfo {author} {\bibfnamefont {T.}~\bibnamefont
  {Mehen}},\ }\bibfield  {title} {\bibinfo {title} {Dilute bose-einstein
  condensate with large scattering length},\ }\href@noop {} {\bibfield
  {journal} {\bibinfo  {journal} {Physical review letters}\ }\textbf {\bibinfo
  {volume} {88}},\ \bibinfo {pages} {040401} (\bibinfo {year}
  {2002})}\BibitemShut {NoStop}%
\bibitem [{\citenamefont {Gammal}\ \emph {et~al.}(2000)\citenamefont {Gammal},
  \citenamefont {Frederico}, \citenamefont {Tomio},\ and\ \citenamefont
  {Abdullaev}}]{16}%
  \BibitemOpen
  \bibfield  {author} {\bibinfo {author} {\bibfnamefont {A.}~\bibnamefont
  {Gammal}}, \bibinfo {author} {\bibfnamefont {T.}~\bibnamefont {Frederico}},
  \bibinfo {author} {\bibfnamefont {L.}~\bibnamefont {Tomio}},\ and\ \bibinfo
  {author} {\bibfnamefont {F.~K.}\ \bibnamefont {Abdullaev}},\ }\bibfield
  {title} {\bibinfo {title} {Stability analysis of the d-dimensional nonlinear
  schr{\"o}dinger equation with trap and two-and three-body interactions},\
  }\href@noop {} {\bibfield  {journal} {\bibinfo  {journal} {Physics Letters
  A}\ }\textbf {\bibinfo {volume} {267}},\ \bibinfo {pages} {305} (\bibinfo
  {year} {2000})}\BibitemShut {NoStop}%
\bibitem [{\citenamefont {Abdullaev}\ \emph {et~al.}(2001)\citenamefont
  {Abdullaev}, \citenamefont {Gammal}, \citenamefont {Tomio},\ and\
  \citenamefont {Frederico}}]{17}%
  \BibitemOpen
  \bibfield  {author} {\bibinfo {author} {\bibfnamefont {F.~K.}\ \bibnamefont
  {Abdullaev}}, \bibinfo {author} {\bibfnamefont {A.}~\bibnamefont {Gammal}},
  \bibinfo {author} {\bibfnamefont {L.}~\bibnamefont {Tomio}},\ and\ \bibinfo
  {author} {\bibfnamefont {T.}~\bibnamefont {Frederico}},\ }\bibfield  {title}
  {\bibinfo {title} {Stability of trapped bose-einstein condensates},\
  }\href@noop {} {\bibfield  {journal} {\bibinfo  {journal} {Physical Review
  A}\ }\textbf {\bibinfo {volume} {63}},\ \bibinfo {pages} {043604} (\bibinfo
  {year} {2001})}\BibitemShut {NoStop}%
\bibitem [{\citenamefont {McClelland}(1995)}]{20}%
  \BibitemOpen
  \bibfield  {author} {\bibinfo {author} {\bibfnamefont {J.~J.}\ \bibnamefont
  {McClelland}},\ }\bibfield  {title} {\bibinfo {title} {Atom-optical
  properties of a standing-wave light field},\ }\href@noop {} {\bibfield
  {journal} {\bibinfo  {journal} {JOSA B}\ }\textbf {\bibinfo {volume} {12}},\
  \bibinfo {pages} {1761} (\bibinfo {year} {1995})}\BibitemShut {NoStop}%
\bibitem [{\citenamefont {McDonald}\ \emph {et~al.}(2014)\citenamefont
  {McDonald}, \citenamefont {Kuhn}, \citenamefont {Hardman}, \citenamefont
  {Bennetts}, \citenamefont {Everitt}, \citenamefont {Altin}, \citenamefont
  {Debs}, \citenamefont {Close},\ and\ \citenamefont {Robins}}]{29}%
  \BibitemOpen
  \bibfield  {author} {\bibinfo {author} {\bibfnamefont {G.~D.}\ \bibnamefont
  {McDonald}}, \bibinfo {author} {\bibfnamefont {C.~C.}\ \bibnamefont {Kuhn}},
  \bibinfo {author} {\bibfnamefont {K.~S.}\ \bibnamefont {Hardman}}, \bibinfo
  {author} {\bibfnamefont {S.}~\bibnamefont {Bennetts}}, \bibinfo {author}
  {\bibfnamefont {P.~J.}\ \bibnamefont {Everitt}}, \bibinfo {author}
  {\bibfnamefont {P.~A.}\ \bibnamefont {Altin}}, \bibinfo {author}
  {\bibfnamefont {J.~E.}\ \bibnamefont {Debs}}, \bibinfo {author}
  {\bibfnamefont {J.~D.}\ \bibnamefont {Close}},\ and\ \bibinfo {author}
  {\bibfnamefont {N.~P.}\ \bibnamefont {Robins}},\ }\bibfield  {title}
  {\bibinfo {title} {Bright solitonic matter-wave interferometer},\ }\href@noop
  {} {\bibfield  {journal} {\bibinfo  {journal} {Physical review letters}\
  }\textbf {\bibinfo {volume} {113}},\ \bibinfo {pages} {013002} (\bibinfo
  {year} {2014})}\BibitemShut {NoStop}%
\bibitem [{\citenamefont {Ketterle}\ and\ \citenamefont {Inouye}(2001)}]{5_21}%
  \BibitemOpen
  \bibfield  {author} {\bibinfo {author} {\bibfnamefont {W.}~\bibnamefont
  {Ketterle}}\ and\ \bibinfo {author} {\bibfnamefont {S.}~\bibnamefont
  {Inouye}},\ }\bibfield  {title} {\bibinfo {title} {Collective enhancement and
  suppression in bose--einstein condensates},\ }\href@noop {} {\bibfield
  {journal} {\bibinfo  {journal} {Comptes Rendus de l'Acad{\'e}mie des
  Sciences-Series IV-Physics}\ }\textbf {\bibinfo {volume} {2}},\ \bibinfo
  {pages} {339} (\bibinfo {year} {2001})}\BibitemShut {NoStop}%
\bibitem [{\citenamefont {Kozuma}\ \emph {et~al.}(1999)\citenamefont {Kozuma},
  \citenamefont {Deng}, \citenamefont {Hagley}, \citenamefont {Wen},
  \citenamefont {Lutwak}, \citenamefont {Helmerson}, \citenamefont {Rolston},\
  and\ \citenamefont {Phillips}}]{5_25}%
  \BibitemOpen
  \bibfield  {author} {\bibinfo {author} {\bibfnamefont {M.}~\bibnamefont
  {Kozuma}}, \bibinfo {author} {\bibfnamefont {L.}~\bibnamefont {Deng}},
  \bibinfo {author} {\bibfnamefont {E.~W.}\ \bibnamefont {Hagley}}, \bibinfo
  {author} {\bibfnamefont {J.}~\bibnamefont {Wen}}, \bibinfo {author}
  {\bibfnamefont {R.}~\bibnamefont {Lutwak}}, \bibinfo {author} {\bibfnamefont
  {K.}~\bibnamefont {Helmerson}}, \bibinfo {author} {\bibfnamefont
  {S.}~\bibnamefont {Rolston}},\ and\ \bibinfo {author} {\bibfnamefont {W.~D.}\
  \bibnamefont {Phillips}},\ }\bibfield  {title} {\bibinfo {title} {Coherent
  splitting of bose-einstein condensed atoms with optically induced bragg
  diffraction},\ }\href@noop {} {\bibfield  {journal} {\bibinfo  {journal}
  {Physical Review Letters}\ }\textbf {\bibinfo {volume} {82}},\ \bibinfo
  {pages} {871} (\bibinfo {year} {1999})}\BibitemShut {NoStop}%
\bibitem [{\citenamefont {Giltner}\ \emph {et~al.}(1995)\citenamefont
  {Giltner}, \citenamefont {McGowan},\ and\ \citenamefont {Lee}}]{5_26}%
  \BibitemOpen
  \bibfield  {author} {\bibinfo {author} {\bibfnamefont {D.~M.}\ \bibnamefont
  {Giltner}}, \bibinfo {author} {\bibfnamefont {R.~W.}\ \bibnamefont
  {McGowan}},\ and\ \bibinfo {author} {\bibfnamefont {S.~A.}\ \bibnamefont
  {Lee}},\ }\bibfield  {title} {\bibinfo {title} {Atom interferometer based on
  bragg scattering from standing light waves},\ }\href@noop {} {\bibfield
  {journal} {\bibinfo  {journal} {Physical review letters}\ }\textbf {\bibinfo
  {volume} {75}},\ \bibinfo {pages} {2638} (\bibinfo {year}
  {1995})}\BibitemShut {NoStop}%
\bibitem [{\citenamefont {Blakie}\ and\ \citenamefont {Ballagh}(2000)}]{5_28}%
  \BibitemOpen
  \bibfield  {author} {\bibinfo {author} {\bibfnamefont {P.~B.}\ \bibnamefont
  {Blakie}}\ and\ \bibinfo {author} {\bibfnamefont {R.~J.}\ \bibnamefont
  {Ballagh}},\ }\bibfield  {title} {\bibinfo {title} {Mean-field treatment of
  bragg scattering from a bose-einstein condensate},\ }\href@noop {} {\bibfield
   {journal} {\bibinfo  {journal} {Journal of Physics B: Atomic, Molecular and
  Optical Physics}\ }\textbf {\bibinfo {volume} {33}},\ \bibinfo {pages} {3961}
  (\bibinfo {year} {2000})}\BibitemShut {NoStop}%
\bibitem [{\citenamefont {Blakie}(2001)}]{5_33}%
  \BibitemOpen
  \bibfield  {author} {\bibinfo {author} {\bibfnamefont {P.~B.}\ \bibnamefont
  {Blakie}},\ }\emph {\bibinfo {title} {Optical Manipulation of Bose-Einstein
  Condensates}},\ \href@noop {} {Ph.D. thesis},\ \bibinfo  {school} {PhD
  thesis, University of Otago, Dunedin, New Zealand} (\bibinfo {year}
  {2001})\BibitemShut {NoStop}%
\bibitem [{\citenamefont {Trippenbach}\ \emph {et~al.}(2000)\citenamefont
  {Trippenbach}, \citenamefont {Band},\ and\ \citenamefont {Julienne}}]{5_29}%
  \BibitemOpen
  \bibfield  {author} {\bibinfo {author} {\bibfnamefont {M.}~\bibnamefont
  {Trippenbach}}, \bibinfo {author} {\bibfnamefont {Y.~B.}\ \bibnamefont
  {Band}},\ and\ \bibinfo {author} {\bibfnamefont {P.~S.}\ \bibnamefont
  {Julienne}},\ }\bibfield  {title} {\bibinfo {title} {Theory of four-wave
  mixing of matter waves from a bose-einstein condensate},\ }\href@noop {}
  {\bibfield  {journal} {\bibinfo  {journal} {Physical Review A}\ }\textbf
  {\bibinfo {volume} {62}},\ \bibinfo {pages} {023608} (\bibinfo {year}
  {2000})}\BibitemShut {NoStop}%
\bibitem [{\citenamefont {Everitt}\ \emph {et~al.}(2017)\citenamefont
  {Everitt}, \citenamefont {Sooriyabandara}, \citenamefont {Guasoni},
  \citenamefont {Wigley}, \citenamefont {Wei}, \citenamefont {McDonald},
  \citenamefont {Hardman}, \citenamefont {Manju}, \citenamefont {Close},
  \citenamefont {Kuhn}, \citenamefont {Szigeti}, \citenamefont {Kivshar},\ and\
  \citenamefont {Robins}}]{34}%
  \BibitemOpen
  \bibfield  {author} {\bibinfo {author} {\bibfnamefont {P.~J.}\ \bibnamefont
  {Everitt}}, \bibinfo {author} {\bibfnamefont {M.~A.}\ \bibnamefont
  {Sooriyabandara}}, \bibinfo {author} {\bibfnamefont {M.}~\bibnamefont
  {Guasoni}}, \bibinfo {author} {\bibfnamefont {P.~B.}\ \bibnamefont {Wigley}},
  \bibinfo {author} {\bibfnamefont {C.~H.}\ \bibnamefont {Wei}}, \bibinfo
  {author} {\bibfnamefont {G.~D.}\ \bibnamefont {McDonald}}, \bibinfo {author}
  {\bibfnamefont {K.~S.}\ \bibnamefont {Hardman}}, \bibinfo {author}
  {\bibfnamefont {P.}~\bibnamefont {Manju}}, \bibinfo {author} {\bibfnamefont
  {J.~D.}\ \bibnamefont {Close}}, \bibinfo {author} {\bibfnamefont {C.~C.~N.}\
  \bibnamefont {Kuhn}}, \bibinfo {author} {\bibfnamefont {S.~S.}\ \bibnamefont
  {Szigeti}}, \bibinfo {author} {\bibfnamefont {Y.~S.}\ \bibnamefont
  {Kivshar}},\ and\ \bibinfo {author} {\bibfnamefont {N.~P.}\ \bibnamefont
  {Robins}},\ }\bibfield  {title} {\bibinfo {title} {Observation of a
  modulational instability in bose-einstein condensates},\ }\href
  {https://doi.org/10.1103/PhysRevA.96.041601} {\bibfield  {journal} {\bibinfo
  {journal} {Phys. Rev. A}\ }\textbf {\bibinfo {volume} {96}},\ \bibinfo
  {pages} {041601} (\bibinfo {year} {2017})}\BibitemShut {NoStop}%
\bibitem [{\citenamefont {Altin}\ \emph {et~al.}(2011)\citenamefont {Altin},
  \citenamefont {Dennis}, \citenamefont {McDonald}, \citenamefont {Doering},
  \citenamefont {Debs}, \citenamefont {Close}, \citenamefont {Savage},\ and\
  \citenamefont {Robins}}]{n6}%
  \BibitemOpen
  \bibfield  {author} {\bibinfo {author} {\bibfnamefont {P.}~\bibnamefont
  {Altin}}, \bibinfo {author} {\bibfnamefont {G.}~\bibnamefont {Dennis}},
  \bibinfo {author} {\bibfnamefont {G.}~\bibnamefont {McDonald}}, \bibinfo
  {author} {\bibfnamefont {D.}~\bibnamefont {Doering}}, \bibinfo {author}
  {\bibfnamefont {J.}~\bibnamefont {Debs}}, \bibinfo {author} {\bibfnamefont
  {J.}~\bibnamefont {Close}}, \bibinfo {author} {\bibfnamefont
  {C.}~\bibnamefont {Savage}},\ and\ \bibinfo {author} {\bibfnamefont
  {N.}~\bibnamefont {Robins}},\ }\bibfield  {title} {\bibinfo {title} {Collapse
  and three-body loss in a 85 rb bose-einstein condensate},\ }\href@noop {}
  {\bibfield  {journal} {\bibinfo  {journal} {Physical Review A}\ }\textbf
  {\bibinfo {volume} {84}},\ \bibinfo {pages} {033632} (\bibinfo {year}
  {2011})}\BibitemShut {NoStop}%
\bibitem [{\citenamefont {Roberts}\ \emph {et~al.}(2000)\citenamefont
  {Roberts}, \citenamefont {Claussen}, \citenamefont {Cornish},\ and\
  \citenamefont {Wieman}}]{II_3}%
  \BibitemOpen
  \bibfield  {author} {\bibinfo {author} {\bibfnamefont {J.~L.}\ \bibnamefont
  {Roberts}}, \bibinfo {author} {\bibfnamefont {N.~R.}\ \bibnamefont
  {Claussen}}, \bibinfo {author} {\bibfnamefont {S.~L.}\ \bibnamefont
  {Cornish}},\ and\ \bibinfo {author} {\bibfnamefont {C.~E.}\ \bibnamefont
  {Wieman}},\ }\bibfield  {title} {\bibinfo {title} {Magnetic field dependence
  of ultracold inelastic collisions near a feshbach resonance},\ }\href@noop {}
  {\bibfield  {journal} {\bibinfo  {journal} {Physical Review Letters}\
  }\textbf {\bibinfo {volume} {85}},\ \bibinfo {pages} {728} (\bibinfo {year}
  {2000})}\BibitemShut {NoStop}%
\bibitem [{\citenamefont {Grimm}\ \emph {et~al.}(2000)\citenamefont {Grimm},
  \citenamefont {Weidem{\"u}ller},\ and\ \citenamefont {Ovchinnikov}}]{37}%
  \BibitemOpen
  \bibfield  {author} {\bibinfo {author} {\bibfnamefont {R.}~\bibnamefont
  {Grimm}}, \bibinfo {author} {\bibfnamefont {M.}~\bibnamefont
  {Weidem{\"u}ller}},\ and\ \bibinfo {author} {\bibfnamefont {Y.~B.}\
  \bibnamefont {Ovchinnikov}},\ }\bibfield  {title} {\bibinfo {title} {Optical
  dipole traps for neutral atoms},\ }in\ \href@noop {} {\emph {\bibinfo
  {booktitle} {Advances in atomic, molecular, and optical physics}}},\
  Vol.~\bibinfo {volume} {42}\ (\bibinfo  {publisher} {Elsevier},\ \bibinfo
  {year} {2000})\ pp.\ \bibinfo {pages} {95--170}\BibitemShut {NoStop}%
\bibitem [{\citenamefont {Gordon}\ and\ \citenamefont {Ashkin}(1980)}]{19}%
  \BibitemOpen
  \bibfield  {author} {\bibinfo {author} {\bibfnamefont {J.}~\bibnamefont
  {Gordon}}\ and\ \bibinfo {author} {\bibfnamefont {A.}~\bibnamefont
  {Ashkin}},\ }\bibfield  {title} {\bibinfo {title} {Motion of atoms in a
  radiation trap},\ }\href@noop {} {\bibfield  {journal} {\bibinfo  {journal}
  {Physical Review A}\ }\textbf {\bibinfo {volume} {21}},\ \bibinfo {pages}
  {1606} (\bibinfo {year} {1980})}\BibitemShut {NoStop}%
\bibitem [{\citenamefont {McGloin}\ \emph {et~al.}(2003)\citenamefont
  {McGloin}, \citenamefont {Spalding}, \citenamefont {Melville}, \citenamefont
  {Sibbett},\ and\ \citenamefont {Dholakia}}]{38}%
  \BibitemOpen
  \bibfield  {author} {\bibinfo {author} {\bibfnamefont {D.}~\bibnamefont
  {McGloin}}, \bibinfo {author} {\bibfnamefont {G.~C.}\ \bibnamefont
  {Spalding}}, \bibinfo {author} {\bibfnamefont {H.}~\bibnamefont {Melville}},
  \bibinfo {author} {\bibfnamefont {W.}~\bibnamefont {Sibbett}},\ and\ \bibinfo
  {author} {\bibfnamefont {K.}~\bibnamefont {Dholakia}},\ }\bibfield  {title}
  {\bibinfo {title} {Applications of spatial light modulators in atom optics},\
  }\href@noop {} {\bibfield  {journal} {\bibinfo  {journal} {Optics Express}\
  }\textbf {\bibinfo {volume} {11}},\ \bibinfo {pages} {158} (\bibinfo {year}
  {2003})}\BibitemShut {NoStop}%
\bibitem [{\citenamefont {Zhu}\ and\ \citenamefont {Wang}(2014)}]{39}%
  \BibitemOpen
  \bibfield  {author} {\bibinfo {author} {\bibfnamefont {L.}~\bibnamefont
  {Zhu}}\ and\ \bibinfo {author} {\bibfnamefont {J.}~\bibnamefont {Wang}},\
  }\bibfield  {title} {\bibinfo {title} {Arbitrary manipulation of spatial
  amplitude and phase using phase-only spatial light modulators},\ }\href@noop
  {} {\bibfield  {journal} {\bibinfo  {journal} {Scientific reports}\ }\textbf
  {\bibinfo {volume} {4}},\ \bibinfo {pages} {7441} (\bibinfo {year}
  {2014})}\BibitemShut {NoStop}%
\bibitem [{\citenamefont {Altin}\ \emph {et~al.}(2012)\citenamefont {Altin}
  \emph {et~al.}}]{5_30}%
  \BibitemOpen
  \bibfield  {author} {\bibinfo {author} {\bibfnamefont {P.~A.}\ \bibnamefont
  {Altin}} \emph {et~al.},\ }\bibfield  {title} {\bibinfo {title} {The role of
  interactions in atom interferometry with bose-condensed atoms},\ }\href@noop
  {} {\  (\bibinfo {year} {2012})}\BibitemShut {NoStop}%
\end{thebibliography}%
\end{document}